\documentclass[lettersize,journal]{IEEEtran}
\usepackage{cite}
\usepackage{amsmath,amssymb,amsfonts}
\usepackage{algorithmic}
\usepackage{graphicx}
\usepackage{textcomp}
\usepackage{xcolor}
\def\BibTeX{{\rm B\kern-.05em{\sc i\kern-.025em b}\kern-.08em
    T\kern-.1667em\lower.7ex\hbox{E}\kern-.125emX}}

\usepackage{mathabx}
\usepackage{array}
\usepackage{mdwmath}
\usepackage{mdwtab}
\usepackage{eqparbox}
\usepackage{caption}
\usepackage{subcaption}
\usepackage{url}
\usepackage{booktabs}
\usepackage{eurosym}
\usepackage{cite}
\usepackage{stackengine}

\usepackage{float}
\usepackage{tikz}
\usepackage{cleveref}
\usepackage{enumitem}

\usepackage{nomencl}
\makenomenclature

\usepackage{etoolbox}
\renewcommand\nomgroup[1]{%
  \item[\bfseries
  \ifstrequal{#1}{A}{Sets}{%
  \ifstrequal{#1}{B}{Parameters}{%
  \ifstrequal{#1}{C}{Variables}{}}}%
]}

\usepackage{color}
\definecolor{R1}{rgb}{0.5,0,0.5}
\definecolor{R2}{rgb}{0,0.75,0.5}
\definecolor{R3}{rgb}{0.75,0.5,0}

\definecolor{blue}{rgb}{0,0,0}
\color{blue}

\begin{document}
\begingroup
\allowdisplaybreaks

\title{Evaluating Offshore Electricity Market Design Considering Endogenous Infrastructure Investments: Zonal or Nodal?}

\author{Michiel Kenis, 
Vladimir Dvorkin,~\IEEEmembership{Member,~IEEE,}
Tim Schittekatte,
Kenneth Bruninx,~\IEEEmembership{Member,~IEEE,}\\
Erik Delarue,~\IEEEmembership{Member,~IEEE,}
Audun Botterud,~\IEEEmembership{Member,~IEEE}

\thanks{The work of M. Kenis is supported by a Ph.D. grant provided by the Flemish Institute for Technological Research (VITO). 

M. Kenis and E. Delarue are with the Division of Applied Mechanics and Energy Conversion (TME), Department of Mechanical Engineering, KU Leuven, Celestijnenlaan 300 (box 2421), 3001, Leuven, Belgium. They are also with the EnergyVille, Thor Park 8310, 3600, Genk, Belgium. M. Kenis is also with the Flemish Institute for Technological Research (VITO), Boeretang 200, B-2400, Mol, Belgium. (e-mail: \{michiel.kenis,erik.delarue\}@kuleuven.be).

V. Dvorkin is with the Department of Electrical Engineering and Computer Science, University of Michigan, Ann Arbor, MI 48109, USA (e-mail: dvorkin@umich.edu).

T. Schittekatte is with the MIT Sloan School of Management, Massachusetts Institute of Technology, Cambridge, MA 02142, USA, and is with the Florence School of Regulation, European University Institute, Firenze, FI 50133, Italy (e-mail: schtim@mit.edu). 

K. Bruninx is with the Faculty of Technology, Policy and Management, TU Delft, 2628 Delft, The Netherlands. He is also with the Division of Applied Mechanics and Energy Conversion (TME), Department of Mechanical Engineering, KU Leuven, Celestijnenlaan 300 (box 2421), 3001, Leuven, Belgium (e-mail: k.bruninx@tudelft.nl).

% E. Delarue are with the Division of Applied Mechanics and Energy Conversion (TME), Department of Mechanical Engineering, KU Leuven, Celestijnenlaan 300 (box 2421), 3001, Leuven, Belgium. He is also with the EnergyVille, Thor Park 8310, 3600, Genk, Belgium (e-mail: erik.delarue@kuleuven.be). 

A. Botterud is with the Laboratory for Information and Decision Systems, Massachusetts Institute of Technology, Cambridge MA 02142, USA (e-mail: audunb@mit.edu). 
}
}

% The paper headers
\markboth{IEEE Transactions on Energy Markets, Policy and Regulation}%
{}

\maketitle

\begin{abstract}
Policy makers are formulating offshore energy infrastructure plans, including wind turbines, electrolyzers, and HVDC transmission lines. An effective market design is crucial to guide cost-efficient investments and dispatch decisions. This paper jointly studies the impact of offshore market design choices on the investment in offshore electrolyzers and HVDC transmission capacity. We present a bilevel model that incorporates investments in offshore energy infrastructure, day-ahead market dispatch, and potential redispatch actions near real-time to ensure transmission constraints are respected. Our findings demonstrate that full nodal pricing, i.e., nodal pricing both onshore and offshore, outperforms the onshore zonal combined with offshore nodal pricing or offshore zonal layouts. While combining onshore zonal with offshore nodal pricing can be considered as a second-best option, it generally diminishes the profitability of offshore wind farms. However, if investment costs of offshore electrolyzers are relatively low, they can serve as catalysts to increase the revenues of the offshore wind farms. This study contributes to the understanding of market designs for highly interconnected offshore power systems, offering insights into the impact of congestion pricing methodologies on investment decisions. Besides, it is useful towards understanding the interaction of offshore loads like electrolyzers with financial support mechanisms for offshore wind farms.
\end{abstract}


\nomenclature[A, 1]{$\mathcal{G}$}{\small Generators}
\nomenclature[A, 2]{$\mathcal{E}$}{\small electrolyzers}
\nomenclature[A, 3]{$\mathcal{N}$}{\small Nodes}
\nomenclature[A, 4]{$\mathcal{L}$}{\small AC lines}
\nomenclature[A, 5]{$\mathcal{H}$}{\small DC lines}
\nomenclature[A, 6]{$\mathcal{H}'$}{\small Cross-border DC lines}
\nomenclature[A, 7]{$\mathcal{T}$}{\small Timesteps}

\nomenclature[B, 01]{$MC_g$}{\small Marginal cost of generator $g$ [\EUR{}/MWh]}
\nomenclature[B, 02]{$U_n$}{ \small Utility of consumption as end product at node $n$  [\EUR{}/MWh]}
\nomenclature[B, 02]{$U_{e,t}^{H2}$}{ \small Utility of consumption at electrolyzer $e$ [\EUR{}/MWh]}
\nomenclature[B, 03]{$C_e$, $C_h$}{ \small Unit investment cost of electrolyzer $e$ or line $h$ [\EUR{}/MW]}
\nomenclature[B, 04]{$\Bar{F}_h^{DC}$}{ \small Maximum capacity of a new DC line $h$ [MW]}
\nomenclature[B, 05]{$\Bar{D}_{n,t}$}{ \small Demand for electricity as end product at node $n$ [MW]}
\nomenclature[B, 06]{$\Bar{D}^{H2}_{e,t}$}{ \small Maximum capacity of electrolyzer $e$  [MW]}
\nomenclature[B, 07]{$LF_{g,t}$}{ \small Load factor of generator $g$ [-]}
\nomenclature[B, 08]{$nPTDF^n_l$,$zPTDF^z_l$}{ \small Nodal and zonal PTDF [-]}
\nomenclature[B, 09]{$\Bar{f}^{AC}_l$}{ \small Capacity of AC line $l$ [MW]}
\nomenclature[B, 10]{$RAM_l$}{ \small Remaining Available Margin of AC line $l$ [MW]}
\nomenclature[B, 11]{$NTC_h$}{ \small Net Transfer Capacity of DC line $h$ [MW]}
\nomenclature[B, 12]{$a$}{ \small Redispatch cost mark-up [-]}
\nomenclature[B, 13]{$I_{h,n}$, $I_{h,z}$}{ \small Connection between DC line $h$ and node $n$ or zone $z$ [-]}
\nomenclature[B, 14]{$I_{l,n}$, $I_{l,z}$}{ \small Connection between AC line $l$ and node $n$ or zone $z$ [-]}

\nomenclature[C, 01]{$TIC$}{ \small Transmission investment cost [\EUR{}]}
\nomenclature[C, 02]{$EIC$}{ \small electrolyzers' investment cost [\EUR{}]}
\nomenclature[C, 03]{$S_t$}{ \small Economic market surplus [\EUR{}]}
\nomenclature[C, 04]{$R_t$}{ \small Redispatch cost [\EUR{}]}
\nomenclature[C, 05]{$\Bar{f}^{DC}_h$}{ \small Capacity of DC line $h$ [MW]}
\nomenclature[C, 06]{$\Bar{d}^{H2}_e$}{ \small Capacity of electrolyzer $e$ [MW]}
\nomenclature[C, 07]{$y_{g,t}$}{ \small Market dispatch of generator $g$ [MW]}
\nomenclature[C, 08]{$d_{n,t}$}{ \small Market dispatch of end consumption at node $n$ [MW]}
\nomenclature[C, 09]{$d^{H2}_{e,t}$}{ \small Market dispatch of consumption at electrolyzer $e$ [MW]}
\nomenclature[C, 10]{$f^{AC}_{l,t}$,$f^{DC}_{h,t}$}{ \small Flow captured by the market on line $l$ or $h$ [MW]}
\nomenclature[C, 11]{$y^{r}_{g,t}$}{ \small Real-time production of generator $g$ [MW]}
\nomenclature[C, 12]{$d^{r}_{n,t}$}{ \small Real-time end consumption at node $n$ [MW]}
\nomenclature[C, 13]{$d^{H2,r}_{e,t}$}{ \small Real-time consumption at electrolyzer $e$ [MW]}
\nomenclature[C, 14]{$y^{U}_{g,t}$,$y^{D}_{g,t}$}{ \small Up/down redispatch of generator $g$ [MW]}
\nomenclature[C, 15]{$d^{U}_{n,t}$,$d^{D}_{n,t}$}{ \small Up/down redispatch of end consumption at node $n$ [MW]}
\nomenclature[C, 16]{$d^{H2,U}_{e,t}$,$d^{H2,D}_{e,t}$}{ \small Up/down redispatch of electrolyzer $e$ [MW]}
\nomenclature[C, 17]{$f^{AC,r}_{l,t}$,$f^{DC,r}_{h,t}$}{ \small Real-time flow on line $l$ or $h$ [MW]}

\normalsize
\printnomenclature

\section{Introduction} \label{sec:introduction}

\IEEEPARstart{T}{he} total installed offshore wind power capacity worldwide amounted to 56 GW in 2021 and is expected to strongly increase in the future \cite{statista}. Besides, other technologies like storage and electrolyzers are expected to be deployed offshore \cite{northseaenergy2022}, partly as offshore energy hubs \cite{north2022}. In these hubs, the wind farms are connected with an interconnector, typically using the high-voltage direct current (HVDC) technology \cite{ENTSOE2021b}.

In addition to government plans, electricity prices guide investments. The market design\footnote{In this paper, we use the term `market design' to refer to methodologies used for pricing and congestion management.}, which differs across regions, determines how offshore electricity wholesale prices are set. Under nodal pricing, the transmission network is incorporated in the market clearing resulting in potentially unique prices per node as used in, e.g., most U.S. electricity markets. In contrast, under zonal pricing, network constraints are simplified in the market clearing and one price per predefined bidding zone is calculated. Bidding zone borders typically align with national borders as in the EU.

Under the status quo in the European zonal day-ahead electricity marktet, offshore energy hubs would be part of the mainland zone they are (administratively) connected with. This means that the power prices in those hubs would equal mainland prices of the country the hub belongs to. However, the EU strategy for the deployment of offshore infrastructure opts for (an) off-shore market zone(s) to integrate these assets in the European electricity market \cite{eucommission2020}. In contrast, along the US East Coast, the natural choice is that offshore nodal pricing will be opted for to integrate future offshore infrastructure into the wholesale electricity market \cite{pfeifenberger}. There are generally four different market design options to integrate offshore hubs as Tab. \ref{tab:marketdesigns} summarizes.

\begin{table}
    \centering
    \caption{Overview of market design options.}
    \label{tab:marketdesigns}
    \begin{tabular}{l  l  l }
    \toprule
        Market design & Onshore & Offshore \\
    \midrule
        Full Nodal Pricing (FNP) & Nodal & Nodal \\
        Offshore Nodal Pricing (ONP) & Zonal & Nodal \\
        Offshore Zonal Pricing (OZP) & Zonal & One offshore zone \\
        Full Zonal Pricing (FZP) & Zonal & Nodes part of onshore zone \\
    \bottomrule
    \end{tabular}
\end{table}

Literature on onshore market design does not apply to an offshore context because (i) the offshore market participants mainly consist of (wind power) generators and (ii) the dominant transmission technology is HVDC as opposed to onshore AC technology. \textcolor{blue}{The former implies that electricity prices are set by either near-zero marginal cost wind farms or by the willingness-to-pay of a very limited set of consumers when the constraints on the exchanges between the onshore and offshore system are binding in the market clearing mechanism. The latter means that flows on HVDC lines, as well as their impact on the AC lines, can be accurately captured in the flow-based market clearing problem via Advanced Hybrid Coupling, in contrast to their AC counterparts. Note, however, that by definition only HVDC connections connecting two market zones are monitored in the market clearing problem. Intra-market zone offshore HVDC lines are not observed by the market, whereas congestion on internal AC lines can be avoided by including them in the calculation of the limitations on cross-market zone exchanges, i.e., the flow-based domains. These properties have a significant impact on the resulting prices, planned production schedule and required redispatch.}

Literature on offshore electricity market design is limited and share a number of simplifying assumptions \cite{THEMA2020, promotion2020, NSCOGI2014}. Firstly, while the market design determines the efficiency of the dispatch of generators, the spatial granularity of the power prices also impacts investment decisions \cite{grimm2023}. The latter is ignored in existing studies on offshore pricing \cite{THEMA2020, promotion2020, NSCOGI2014}. Secondly, flow-based market coupling (FBMC) is the target methodology in Europe's zonal electricity markets to calculate and allocate AC transmission capacity in wholesale electricity markets while current studies typically consider a `Net Transfer Capacities'-methodology \cite{schonheit2021guide, hardy2023}. Inclusion of DC transmission capacity under FBMC requires, due to operational transmission characteristics, an adjusted flow-based methodology to calculate and allocate both AC and DC transmission capacity called Advanced Hybrid Coupling \cite{muller2017}. Thirdly, novel technologies like offshore electrolyzers are not considered in current studies on the offshore pricing methodology except in Lüth et al. \cite{luth2022} who show that offshore bidding zones facilitate offshore hydrogen production. With many projects ongoing to install such technologies offshore, it is important to consider them in the analysis because they serve as the only, or at least a very important, offshore electricity demand \cite{northseaenergy2022}. Hence, electrolyzers have the potential to have a large impact on offshore prices \cite{green2011}. 

This paper addresses the gaps identified above by proposing an optimization model that considers all these three elements. \textcolor{blue}{Specifically, we present a bilevel optimization model that captures four temporal stages: (i) offshore HVDC transmission investment, (ii) offshore electrolyzer investment, (iii) day-ahead market clearing, and (iv) redispatch actions.} We assume a fixed capacity target for offshore wind generation investment as these decisions are typically not market-driven. Our results reveal that ONP and OZP can serve as valuable alternatives to FZP. However, these schemes come at the cost of a lower profitability for existing offshore wind farms due to a lower average offshore price. The latter benefits the business case for potential offshore electricity consumers like, e.g., offshore electrolyzers, which can contribute to decrease the welfare gap between ideal FNP, on the one hand, and ONP and OZP on the other hand. As a result, policy makers should carefully re-consider support mechanisms for, e.g., offshore wind power under ONP and OZP.

The paper is structured as follows. Section \ref{sec:literature} discusses the existing literature. Section \ref{sec:model} presents the optimization model, while Section \ref{sec:numericalillustration} demonstrates the model by applying it to a stylized electricity grid. Finally, Section \ref{sec:discussion} discusses the implications of our findings and Section \ref{sec:conclusion} concludes. 

\section{Literature} \label{sec:literature}

There are two main literature streams that relate to this paper: literature on (i) existing optimization models that capture both the investments in and the operation of electricity grids, and (ii) offshore electricity market design with a focus on the spatial granularity of wholesale electricity prices.

\subsection{Capacity expansion models}

\begin{figure*}[h]
    \centering
    \includegraphics[width=0.9\textwidth]{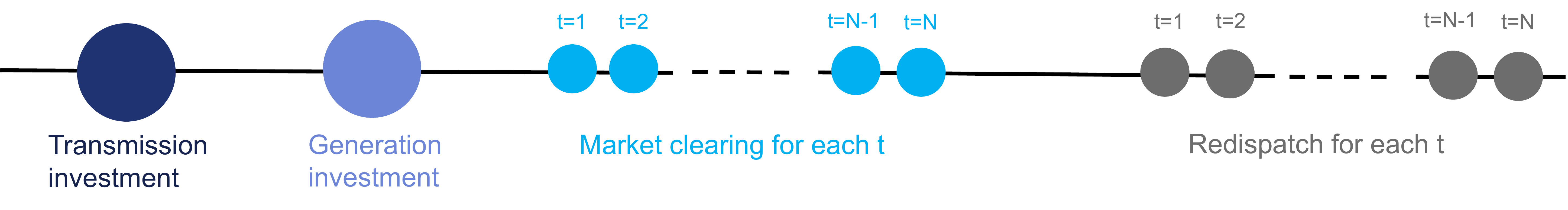}
    \caption{Typical composition of optimization models in literature that capture transmission and generation investments as well as the operation of the considered electricity grid.}
    \label{fig:literaturestructure}
\end{figure*}

With regards to onshore electricity grids, the existing literature has proposed optimization models that encompass transmission and generation investments, as well as the operation of the electricity grid. Figure \ref{fig:literaturestructure} shows the typical composition of these models capturing four stages: transmission investment, generation investment, market clearings and redispatch actions. The structure of these optimization models varies in the literature based on underlying assumptions. The hierarchical or non-hierarchical relationship between agents, each with distinct objectives and decision variables, categorizes such models. Many papers assume that investors strategically consider the impact of their investments on the power system's operation, resulting in hierarchical models \cite{grimm2016, kleinert2019, egerer2021}. Non-hierarchical models, on the other hand, do not incorporate strategic anticipation \cite{hardy2023}.

These approaches are, however, not suitable to analyze offshore power systems, especially when connected to zonal electricity markets (onshore) subject to FBMC as in Europe. FBMC considers the limited capacity of critical intra-zonal or inter-zonal AC lines as a constraint in the market clearing algorithm in a simplified way \cite{van2016flow}. Specifically, the market representation of the flows on AC transmission network elements is capped by their Remaining Available Margins (RAMs), parameters that TSOs provide. The methodology is being rolled out in most European countries \cite{schonheit2021guide}. To consider (offshore) DC lines, an adjusted flow-based market clearing algorithm is required because the flow on DC lines depends on how the AC/DC converter is operated. Two options are being considered: Standard Hybrid Coupling (SHC -- which is currently used) and Advanced Hybrid Coupling (AHC -- target methodology) \cite{muller2017}. AHC, in contrast to SHC, considers the impact of DC flows on the market representation of AC flows. We refer the reader to Schonheit et al. \cite{schonheit2021guide} for an exhaustive guide on the functioning on FBMC, and to Muller et al. \cite{muller2017} for the compatibility with DC lines.

Lete et al. \cite{lete2022} includes FBMC in capacity expansion, but they do not consider DC lines. To the best of the authors' knowledge, our paper is the first to consider both AC and DC lines in a capacity expansion and operation market model by applying AHC. Moreover, an additional novelty is that we optimize investments in both HVDC transmission capacity and offshore electrolysers within the same model.

\subsection{Offshore electricity market design}

The literature on integrating offshore energy infrastructure into existing onshore zonal electricity markets has gained significant attention in the EU after the European Commission highlighted its “strategy to harness the potential of offshore renewable energy for a climate neutral future" \cite{eucommission2020}. Several studies suggest creating offshore bidding zones to enable efficient flows of renewable energy to high-cost areas, reduce redispatch costs, and enhance operational security \cite{THEMA2020, promotion2020, NSCOGI2014, klitzing2020, tosatto2022}. Tosatto et al. \cite{tosatto2022} incorporate FBMC in their models and demonstrate that under offshore bidding zones, offshore electricity prices would decrease while congestion rents would increase. However, none of the aforementioned studies endogenously model how investments in generation and/or transmission assets are impacted by the different price signals under alternative market designs for congestion management.

Hardy et al. \cite{hardy2023} are the first to present a capacity expansion and operational market model within the context of offshore electricity market design. They show that investors should assume that offshore bidding zones will be implemented because the resulting power system (topology as well as generating assets) also performs well under changing pricing methodologies. There are three key differences between their work and our paper. First, in contrast to \cite{hardy2023}, we assume that investors in transmission capacity strategically anticipate the impact on the market clearing and redispatch actions. Second, Hardy et al. \cite{hardy2023} do not consider offshore electrolysers. Lastly, their model neglects FBMC when assessing the performance of offshore bidding zones.

\section{Model} \label{sec:model}

This section presents the high-level structure of the model that is first organized as a quad-level problem, then reformulated to a bilevel problem. \textcolor{blue}{First, we list the assumptions that underpin our model.} Next, we present the mathematical formulation of the bilevel model. Finally, we provide a solution strategy.

\subsection{Assumptions}
\textcolor{blue}{There are four stages and we assume these stages occur sequentially: HVDC transmission investment (by the TSO), electrolyzer investment (by a merchant investor), market clearing (by the market operator), and redispatch (by the TSO) (Fig. 1). Each level anticipates the impact of its decision on later stages, while taking the decision variables of all higher levels as given. We assume perfect foresight in all stages.}

\textcolor{blue}{We assume a fixed capacity for offshore wind generation as such capacity decisions are typically not market-driven, but prescribed in tenders. Similarly, we assume an existing fleet of conventional generators connected to the onshore nodes. Investments are onshore assets is not considered.}

\textcolor{blue}{As we seek to compare the performance of \textit{optimal} system configurations under different market designs, we consider the HVDC capacity as a variable. Fixing the HVDC capacity under different market designs may lead to, e.g., over- or underestimating the benefits for FNP, depending on the grid topology and capacity of the HVDC elements. In the HVDC investment stage, we assume that there is no existing HVDC transmission capacity. We do not consider the lumpiness of network investments and only investments between predefined nodes are allowed (see Section \ref{sec:setup}). Electrolyzer capacity 
is limited to a predefined upper bound per offshore node.}

\textcolor{blue}{The market clearing can adopt four pricing methodologies: FNP, ONP, OZP and FZP. There is perfect competition between participants in the wholesale electricity market. Only thermal generation units face a non-zero variable cost $MC_g$. The marginal utility of demand for electricity to produce hydrogen depends on the hydrogen price and transport cost, which are assumed to be fixed. Under zonal pricing, flow-based market coupling and Advanced Hybrid Coupling are considered. We assume that all AC lines are critical and, hence, the transmission limits of all AC lines are (albeit imperfectly) considered in the market algorithm. We refer the reader to Schönheit et al. \cite{schonheit2021guide} for an exhaustive guide on flow-based market coupling.}

\textcolor{blue}{Redispatch actions only serve to adjust the generators' dispatch in case their output as cleared in the day-ahead market would lead to infeasible power flows. This is only required if the day-ahead market does not accurately capture the transmission constraints, i.e., under ONP, OZP an FZP. We assume cost-based redispatch.}

\subsubsection{The quad-level problem}

\begin{figure}
    \centering
    \includegraphics[width=0.5\textwidth]{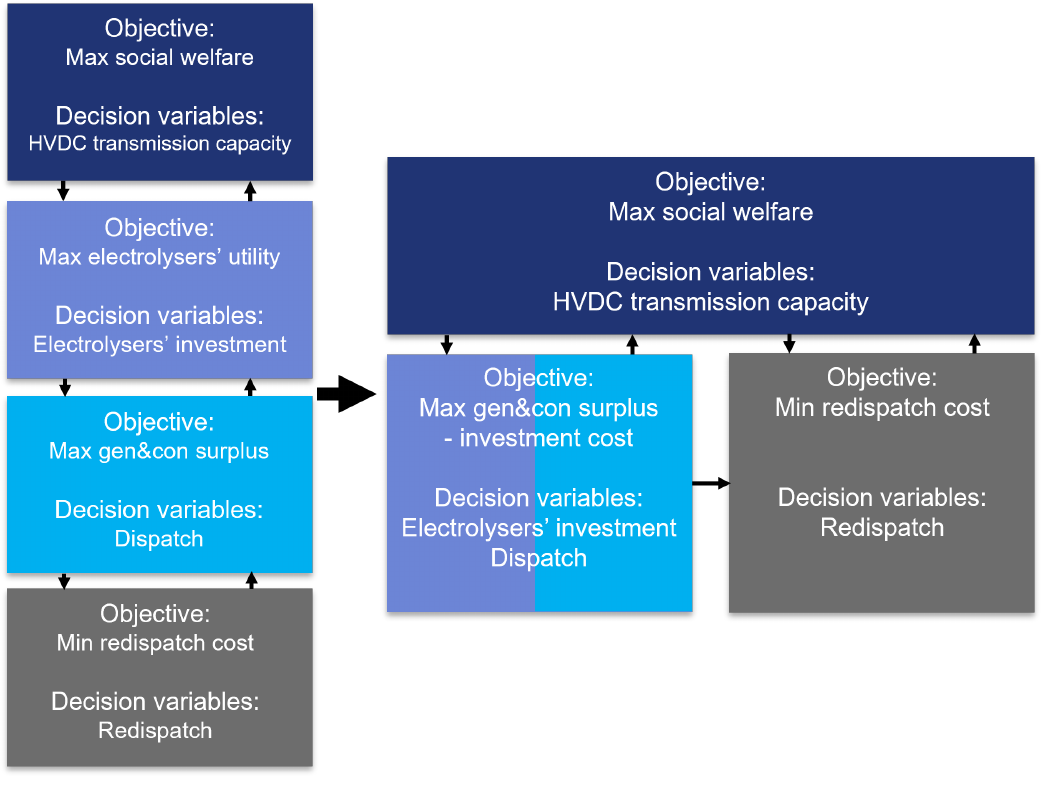}
    \caption{From a quad-level problem to a bilevel problem by assuming perfect competition between generators. The colors map to the different stages in Fig. \ref{fig:literaturestructure}.}
    \label{fig:structure}
\end{figure}

Under the assumptions above, the resulting decision problem can be represented as a quad-level problem (Fig. \ref{fig:structure}). The first level, as shown in the left panel of Fig. \ref{fig:structure}, maximises the overall social welfare while deciding on the HVDC transmission capacity. The second level maximises the net utility of newly added electrolyzers while deciding on the eletrolysers' capacities. The third level maximises the surplus from the day-ahead market with the generators' dispatch (and possible load shedding) as decision variable. Finally, the fourth level minimises redispatch costs including cost mark-ups assuming cost-based redispatch. 

\subsubsection{From a quad-level to a bilevel problem}

We transform the quad-level problem to a bilevel problem in two steps as Fig. \ref{fig:structure} visualises. \textcolor{blue}{The assumption of perfect competition between participants in the wholesale electricity market allows merging the second (electrolyzers' investments) and third (market clearing) level optimization problems into one optimization problem.} The merged problem maximizes the surplus from the wholesale market less the investment cost of electrolyzers. The equivalence has been shown in literature for the case of generators' investments but is equivalent for electrolyzers' investments \cite{grimm2013}. 

Secondly, we note that the decision variables of the fourth level (redispatch) are independent from the second (electrolyzers' investments) or third (market clearing) level optimization problems, but not from the first (transmission investment) level. As a consequence, the redispatch problem can be considered to be happening on the same hierarchical level as the combined electrolyzer investment \& market clearing problem. This is similar to the approach of Pandzic et al. \cite{pandzic2021} in the context of reserve capacity markets and balancing markets.

\subsection{A bilevel problem}\label{sec:math}

We present the mathematical equations of the bilevel problem consisting of two lower level problems. The lower level representing the combined electrolyzer investment \& market clearing problem comes in two forms: either considering full nodal pricing or full zonal pricing. The cases of ONP and OZP are special cases of FZP in which each offshore node or all offshore nodes together represent (a) separate market zone(s). 

\subsubsection{Upper Level: transmission investment}

The decision variables are the HVDC transmission capacity $\mathbf{\bar{f}^{DC}_{h}}$ of each candidate line $h \in \mathcal{H}$ and the transmission investment cost $\mathbf{TIC}$. The decision variables are written in a bold font. We assume no existing HVDC transmission capacity. 

\begin{subequations}\label{eq:UL}
\begin{align}
& \text{max}\quad  \sum_{t \in \mathcal{T}} \Big[ S_t - R_t \Big] - EIC - \mathbf{TIC}  \label{eq:UL_obj}  \\
& \text{s.t.}\nonumber\\
& S_t = \sum_{n \in \mathcal{N}} U_n  d^{}_{n,t} + \sum_{e \in \mathcal{E}} U^{H2}_{e,t}  d^{H2}_{e,t} \nonumber\\ &- \sum_{g \in \mathcal{G}} MC_g  y^{}_{g,t} &  \forall t \label{eq:UL_con1}\\
&R_t = \sum_{n \in \mathcal{N}} U_n \cdot \big[ d_{n,t} - d^{r}_{n,t} \big] + \sum_{e \in \mathcal{E}} U^{H2}_{e,t} \cdot   \nonumber \\ & \big[ d^{H2}_{e,t} -d^{H2,r}_{e,t} \big] + \sum_{g \in \mathcal{G}} MC_g \cdot \big[ y_{g,t}^{r} - y_{g,t} \big]  &  \forall t \label{eq:UL_con2}\\
& EIC = \sum_{e \in \mathcal{E}} C_e  \bar{d}^{H2}_{e}  & \forall g, \forall t  \label{eq:UL_con3}\\
&\mathbf{TIC} = \sum_{h \in \mathcal{H}'} C_h  \mathbf{\bar{f}^{DC}_h}  &  \forall n, \forall t  \label{eq:UL_con4}\\
&0\leq \mathbf{\bar{f}^{DC}_{h}} \leq \bar{F}^{DC}_h &  \forall h \label{eq:UL_con5}
\end{align}
\end{subequations}

Objective (\ref{eq:UL_obj}) maximises the overall social welfare. This is reasonable as a transmission investment is typically a strongly regulated process \cite{joskow2005}. Social welfare consists of the economic surplus of the wholesale electricity market $S_t$ less redispatch costs $R_t$, electrolyzer investment cost $EIC$ and transmission investment cost $\mathbf{TIC}$. Eqs. (\ref{eq:UL_con1})-(\ref{eq:UL_con4}) define each of the terms. Equation (\ref{eq:UL_con5}) poses constraints to the new transmission capacity. \textcolor{blue}{Investments in transmission capacity only occur if the investment cost is outweighed by the operational benefits (higher welfare and/or lower redispatch costs).}

The economic surplus of the wholesale electricity market $S_t$ equals the utility associated with consuming electricity less the operational generation costs. The former splits into the utility of demand for electricity as end-product $U_n \cdot d_{n,t}$ and demand for electricity as a source for electrolysis $U^{H2}_{e,t} \cdot d^{H2}_{e,t}$. The marginal utility of demand for electricity to produce hydrogen depends on the hydrogen price and transport cost. We further assume that only thermal generation units face a non-zero variable cost $MC_g$. 

The redispatch cost $R_t$ is the net cost for a TSO to adjust the dispatch schedule corresponding the day-ahead market to prevent line overloading. Cost-based redispatch implies that a generator, requested by the TSO to produce more than scheduled by the market operator, is paid its marginal cost $MC_g$ by the TSO while a generator, requested by the TSO to produce less than scheduled by the market operator, pays back its avoided marginal cost to the TSO but keeps its profit margin. A similar concept holds for consumers: if the covered demand is lower than scheduled by the market operator, consumers are paid their lost utility $U_n$ or $U^{H2}_{e,t}$, and vice versa. Finally, the electrolyzers' investment cost $EIC$ and transmission investment cost $\mathbf{TIC}$ scale linearly with the installed capacity.

\subsubsection{Lower Level 1: electrolyzers' investment \& market clearing}

Problem (\ref{eq:LL1}) presents a first part of the first lower level problem in mathematical form. Eqs. (\ref{eq:LL1_nodal}), in case of nodal pricing, and Eqs. (\ref{eq:LL1_zonal}), in case of zonal pricing, present the remaining constraints. The decision variables are the new capacity of electrolyzers $\mathbf{\bar{d}^{H2}_{e}}$ as well as the dispatch of all generators $\mathbf{y_{g,t}}$. Other decision variables are the covered demand for electricity as end-product $\mathbf{d_{n,t}}$ and electricity for electrolysis $\mathbf{d^{H2}_{e,t}}$ as well as the power flows on all AC lines $\mathbf{f^{AC}_{l,t}}$ and DC-lines $\mathbf{f^{DC}_{h,l}}$.

\begin{subequations}\label{eq:LL1}
\begin{align}
& \text{max}\quad \sum_{t \in \mathcal{T}} \mathbf{S_t} - \mathbf{EIC}  \label{eq:LL1_obj} \\
& \text{s.t.}\quad \nonumber\\
& \text{Eq. (}\ref{eq:UL_con1}\text{), Eq. (}\ref{eq:UL_con3}\text{)} \label{eq:LL1_con1}\\
&0 \leq \mathbf{\bar{d}^{H2}_{e}} \leq \bar{D}^{H2}_{e}  &  \forall e \label{eq:LL1_con4}\\
&0 \leq \mathbf{y_{g,t}} \leq LF_{g,t} \bar{y}_g  & \forall g, \forall t  \label{eq:LL1_con5}\\
&0 \leq \mathbf{d_{n,t}} \leq \bar{D}_{n,t}  &  \forall n, \forall t  \label{eq:LL1_con8}\\
&0 \leq \mathbf{d^{H2}_{e,t}} \leq \bar{d}^{H2}_{e} &  \forall e, \forall t \label{eq:LL1_con9}
\end{align}
\end{subequations}

Objective (\ref{eq:LL1_obj}) maximises the surplus from the wholesale market $\mathbf{S_t}$ less the electrolyzers' investment cost $\mathbf{EIC}$. Equation (\ref{eq:LL1_con1}) defines both terms in the objective. Equation (\ref{eq:LL1_con4}) puts limitations on the new capacity of offshore electrolyzers. Equations (\ref{eq:LL1_con5})-(\ref{eq:LL1_con9}) limit the dispatch of generators and consumers to the available power capacities. 

\begin{subequations}\label{eq:LL1_nodal}
\begin{align}
& \textit{In case of nodal pricing:} \nonumber \\
& \mathbf{f^{AC}_{l,t}} = \sum_{n \in \mathcal{N}} nPTDF^{n}_{l} \Big[\sum_{g \in \mathcal{G(N)}} \mathbf{y_{g,t}} -\mathbf{d_{n,t}} \nonumber \\ & - \sum_{e \in \mathcal{E}(n)} \mathbf{d^{H2}_{e,t}} - \sum_{h \in \mathcal{H}} \mathbf{f^{DC}_{h,t}} \cdot I_{h,n} \Big] & \forall l, \forall t \label{eq:LL1_nodal_con1}\\
&|\mathbf{f^{AC}_{l,t}}|  \leq \bar{f}^{AC}_l    &  \forall l, \forall t \label{eq:LL1_nodal_con2}\\ 
&|\mathbf{f^{DC}_{h,t}}| \leq \bar{f}^{DC}_h &  \forall h, \forall t \label{eq:LL1_nodal_con3}\\
&\sum_{g \in \mathcal{G}(n)} \mathbf{y_{g,t}} - \mathbf{d_{n,t}} - \sum_{e \in \mathcal{E}(n)} \mathbf{d^{H2}_{e,t}} \nonumber  \\ &- \sum_{h \in \mathcal{H}} \mathbf{f^{DC}_{h,t}} \cdot I_{h,n} -\sum_{l \in \mathcal{L}} \mathbf{f^{AC}_{l,t}} \cdot I_{l,n} = 0 & \forall n, \forall t \label{eq:LL1_nodal_con4}
\end{align}
\end{subequations}

Equation (\ref{eq:LL1_nodal_con1}) defines the flow on each AC line using nodal PTDFs. Ignoring the DC approximation, this is an exact projection of the physical flow as opposed to the case of zonal pricing. The final term considers the injection of the DC grid into the AC grid. $I_{h,n}$ equals 1 (or -1) if the flow on DC-line $h$ goes outward (or inward) node $n$, and is otherwise 0. Eqs. (\ref{eq:LL1_nodal_con2}) and (\ref{eq:LL1_nodal_con3}) limits the flow on all AC and DC lines by their physical capacities. Finally, Eq. (\ref{eq:LL1_nodal_con4}) presents the nodal power balance.

\begin{subequations}\label{eq:LL1_zonal}
\begin{align}
& \textit{In case of zonal pricing:} \nonumber \\
& \mathbf{f^{AC}_{l,t}} = \sum_{z \in \mathcal{Z}} zPTDF^{z}_{l} \Big[\sum_{g \in \mathcal{G}(z)} \mathbf{y_{g,t}} - \sum_{n \in \mathcal{N}(z)} \mathbf{d_{n,t}}  - \sum_{e \in \mathcal{E}(z)} \mathbf{d^{H2}_{e,t}} \nonumber \\ & - \sum_{h \in \mathcal{H}} \mathbf{f^{DC}_{h,t}} I_{h,z} \Big] + \sum_{h \in \mathcal{H}'} \mathbf{f^{DC}_{h,t}} \Big[ nPTDF^{n(h-)}_l \cdot I_{h,n(h-)} \nonumber \\
& + nPTDF^{n(-h)}_l \cdot I_{h,n(-h)} \Big] \hspace{29.5mm} \forall l, \forall t \label{eq:LL1_zonal_con1}\\
&|\mathbf{f^{AC}_{l,t}}|  \leq RAM_l \hspace{48mm}   \forall l, \forall t \label{eq:LL1_zonal_con2}\\ 
&|\mathbf{f^{DC}_{h,t}}| \leq NTC_h \hspace{39mm}   \forall h \in \mathcal{H}', \forall t \label{eq:LL1_zonal_con3}\\
&\sum_{g \in \mathcal{G}(z)} \mathbf{y_{g,t}} - \sum_{n \in \mathcal{Z}} \mathbf{d_{n,t}} - \sum_{e \in \mathcal{E}(z)} \mathbf{d^{H2}_{e,t}} \nonumber \\ & - \sum_{h \in \mathcal{H}} \mathbf{f^{DC}_{h,t}} \cdot I_{h,z} -\sum_{l \in \mathcal{L}} \mathbf{f^{AC}_{l,t}} \cdot I_{l,z} = 0  \hspace{15.5mm} \forall z, \forall t \label{eq:LL1_zonal_con4}
\end{align}
\end{subequations}

Equation (\ref{eq:LL1_zonal_con1}) defines the flow on each AC line using zonal PTDFs\footnote{Following flow-based market coupling, only the flow on critical AC lines should be monitored and capped. We assume here that all AC lines are critical.}. Zonal PTDFs are parameters, set by TSOs, that are predictive and imperfect by definition as they aggregate nodal information. As a consequence, $\mathbf{f^{AC}_l}$ does not capture the physical power flow but rather an imperfect heuristic, hereafter referred to as the commercial flow. The impact of the power flow on a DC line is captured following Advanced Hybrid Coupling \cite{muller2017}. The flow on an AC line $l$ as a result of a flow on a DC line $h$ is calculated with $nPTDF^{n(h-)}_l$ and $nPTDF^{n(-h)}_l$ with $n(h-)$ and $n(-h)$ the starting/ending end of DC-line $h$. $I_{h,n(h-)}$ and $I_{h,n(-h)}$ equal 1 (or -1) if the flow on DC-line $h$ is defined as flowing into (or away from) node $n(h-)$ or $n(-h)$, and is zero otherwise. Equation (\ref{eq:LL1_zonal_con2}) limits the commercial power flow on each AC line by the Remaining Available Margin (RAM), i.e., the part of the physical transmission capacity that can be traded in the wholesale market \cite{schonheit2021guide}. Equation (\ref{eq:LL1_zonal_con3}) limits the flow on each cross-border DC line by the Net Transfer Capacity (NTC). Finally, Equation (\ref{eq:LL1_zonal_con4}) presents the zonal power balance.

\subsubsection{Lower Level 2: redispatch}

The decision variables are the final dispatch, after redispatch, of all generators $\mathbf{y_{g,t}^{r}}$ as well as the final covered demand for electricity as end-product $\mathbf{d_{n,t}^{r}}$ and electricity for electrolysis $\mathbf{d^{H2,r}_{e,t}}$. We also consider explicit variables for the upward and downward adjustment to the wholesale market dispatch $\mathbf{y^{U}_{g,t}}$, $\mathbf{y^{D}_{g,t}}$, $\mathbf{d^{U}_{n,t}}$, $\mathbf{d^{D}_{n,t}}$, $\mathbf{d^{H2,U}_{e,t}}$ and $\mathbf{d^{H2,D}_{e,t}}$. Finally, $\mathbf{f^{AC,r}_{l,t}}$ and $\mathbf{f^{DC,r}_{h,l}}$ are the physical AC and DC power flow after redispatch.

\begin{subequations}\label{eq:LL2}
\begin{align}
\text{min}\quad & \sum_{t \in \mathcal{T}} \sum_{g \in \mathcal{G}} \big[ MC_g + a \big] \cdot \mathbf{y^{U}_{g,t}} + \big[ -MC_g + a \big] \cdot \mathbf{y^{D}_{g,t}} \nonumber \\
&+ \sum_{e \in \mathcal{E}} \big[ -U^{H2}_{e,t} + a \big] \cdot \mathbf{d^{H2,U}_{e,t}} + \big[ U^{H2}_{e,t} + a \big] \cdot \mathbf{d^{H2,D}_{e,t}} \nonumber \\ & + \sum_{n \in \mathcal{N}} \big[ -U_n + a \big] \cdot \mathbf{d^{U}_{n,t}} + \big[ U_n + a \big] \cdot \mathbf{d^{D}_{n,t}} \label{eq:LL2_objective} 
\end{align}
\begin{align}
& \text{s.t.}\nonumber\\
&\mathbf{y_{g,t}^{r}} = y_{g,t}+\mathbf{y^{U}_{g,t}}-\mathbf{y^{D}_{g,t}} &  \forall g, \forall t \label{eq:LL2_con1}\\
&\mathbf{d_{n,t}^{r}} = d_{n,t}+\mathbf{d^{U}_{n,t}}-\mathbf{d^{D}_{n,t}} & \forall n, \forall t \label{eq:LL2_con4}\\
&\mathbf{d_{e,t}^{H2,r}} = d^{H2}_{e,t}+\mathbf{d^{H2,U}_{e,t}}-\mathbf{d^{H2,D}_{e,t}} & \forall e, \forall t \label{eq:LL2_con5}\\
&0 \leq \mathbf{y^{U}_{g,t}} \leq  LF_{g,t} \cdot \bar{y}_g - y_{g,t} & \forall g, \forall t \label{eq:LL2_con6}\\
&0 \leq \mathbf{y^{D}_{g,t}} \leq y_{g,t} & \forall g, \forall t \label{eq:LL2_con7}\\
&0 \leq \mathbf{d^{U}_{n,t}} \leq \bar{D}_n-d_{n,t} & \forall n, \forall t \label{eq:LL2_con12}\\
&0 \leq \mathbf{d^{D}_{n,t}} \leq d_{n,t} & \forall n, \forall t \label{eq:LL2_con13}\\
&0 \leq \mathbf{d^{H2,U}_{e,t}} \leq \bar{d}^{H2}_e-d^{H2}_{e,t} & \forall e, \forall t \label{eq:LL2_con14}\\
&0 \leq \mathbf{d^{H2,D}_{e,t}} \leq d^{H2}_{e,t} & \forall e, \forall t \label{eq:LL2_con15} \\
& \mathbf{f^{AC,r}_{l,t}} = \sum_{n \in \mathcal{N}} nPTDF^{n}_{l} \Big[\sum_{g \in \mathcal{G(N)}} \mathbf{y^{r}_{g,t}} \nonumber \\ & -\mathbf{d^{r}_{n,t}} - \sum_{e \in \mathcal{E}(n)} \mathbf{d^{H2,r}_{e,t}} - \sum_{h \in \mathcal{H}} \mathbf{f^{DC,r}_{h,t}} \cdot I_{h,n} \Big] &  \forall l, \forall t \label{eq:LL2_con18}\\
& |\mathbf{f^{AC,r}_{l,t}}| \leq \bar{f}^{AC}_l & \forall l, \forall t \label{eq:LL2_con19} \\
& |\mathbf{f^{DC,r}_{h,t}}| \leq \bar{f}^{DC}_h &\forall h, \forall t \label{eq:LL2_con20} \\
&\sum_{g \in \mathcal{G}(n)} \mathbf{y^{r}_{g,t}} - \mathbf{d^{r}_{n,t}} - \sum_{e \in \mathcal{E}(n)} \mathbf{d^{H2,r}_{e,t}} - \nonumber \\ & \sum_{h \in \mathcal{H}} \mathbf{f^{DC,r}_{h,t}} \cdot I_{h,n} -\sum_{l \in \mathcal{L}} \mathbf{f^{AC,r}_{l,t}} \cdot I_{l,n} = 0 & \forall n, \forall t \label{eq:LL2_con21}
\end{align}
\end{subequations}

Objective (\ref{eq:LL2_objective}) minimises the redispatch cost considering a cost mark-up $a$ to avoid redispatch actions \textcolor{blue}{when the outcome of the day-ahead market does not lead to overloading of network elements}. This differs from existing papers that do not consider a mark-up \cite{grimm2016, kleinert2019, egerer2021}. Indeed, the aim of the optimisation problem is to determine the required adjustments to the market outcome only to avoid line overloadings, not to correct a possibly inefficient dispatch from the market. This is in line with real world redispatch actions in zonal wholesale markets \cite{weibezahl2017}. Therefore, by imposing cost mark-ups we avoid redispatch when there is no congestion. \textcolor{blue}{The redispatch cost $R_t$, and thus social welfare, in the upper level problem is the \textit{true} redispatch costs, i.e., without cost mark-up $a$.}

Eqs. (\ref{eq:LL2_con1})-(\ref{eq:LL2_con5}) define the final dispatch, after redispatch, of all generators $\mathbf{y_{g,t}^{r}}$ as well as the final covered demand for electricity as end-product $\mathbf{d_{n,t}^{r}}$ and electricity for electrolysis $\mathbf{d^{H2,r}_{e,t}}$. Eqs. (\ref{eq:LL2_con6})-(\ref{eq:LL2_con15}) provide limits to the upward or downward adjustment taking the market outcome (from lower level 1) as given. Finally, similar to Eqs. (\ref{eq:LL1_nodal_con1})-(\ref{eq:LL1_nodal_con4}), Eqs. (\ref{eq:LL2_con18})-(\ref{eq:LL2_con21}) present the definition of the power flow on each AC line, the limitations to the AC and DC power flows, and the nodal power balance in real-time, i.e., after redispatch. 

\subsection{Solution Approach} \label{sec:solutionapproach}

We follow the approach of Kazempour and Conejo \cite{kazempour2012}, and apply the Benders' decomposition to a bilevel problem, considering $\Bar{f}^{DC}_h$ as the complicating variable. The approach works as follows:
\begin{enumerate}
    \item Given fixed HVDC transmission investment decisions, \textcolor{blue}{we solve two single-level reformulations of the bilevel problem. Specifically, after fixing the complicating variable, we first solve the mixed-integer reformulation of the simplified MPEC which now only has lower-level decision variables. Subsequently, we solve a continuous yet nonlinear version of the same problem, now formulated using the strong duality theorem, wherein the variables causing bilinear terms are fixed to the values derived from the solution of the first MPEC. This, in turn, allows the extraction of sensitivities of welfare w.r.t. HVDC investments — which is the main motivation behind solving two MPECs at every iteration, as in \cite{kazempour2012}.}
    \item The sensitivities obtained in step 1 allow for formulating a so-called Benders' master problem whose solution provides updated HVDC investment decisions.
    \item Step 1-2 are repeated until no more economic welfare improvement can be achieved. 
\end{enumerate}

\textcolor{blue}{Our framework resembles that of Kazempour and Conejo \cite{kazempour2012}, with some minor differences that do not affect the solution strategy. The operational decisions in our problem are not optimized by the upper level investor, as in \cite{kazempour2012}, but rather by a system operator. Another distinction lies in our aim to maximize societal welfare as opposed to minimizing negative profit and the absence of scenarios due to our deterministic framework. While our model is distinct in these ways, it does not alter the solution strategy proposed in \cite{kazempour2012}.}

\section{Case Study} \label{sec:numericalillustration}

\subsection{Set-up} \label{sec:setup}

Figure \ref{fig:network} presents the network layout. We consider three mainland zones, each consisting of three nodes, and three offshore nodes. We consider 30 time steps with varying consumption levels and load factors of offshore wind. The size of the case study is similar to related literature \cite{grimm2016, kleinert2019, egerer2021}. Appendix A presents the assumed values of all parameters in the algorithm. The algorithm, applying Benders' decomposition, converges after less than 20 iterations and a total run time of less than $10^4$ seconds for all cases considering a tolerance of $10^{-1}$.

\begin{figure}
    \centering
    \includegraphics[width = 0.5\textwidth]{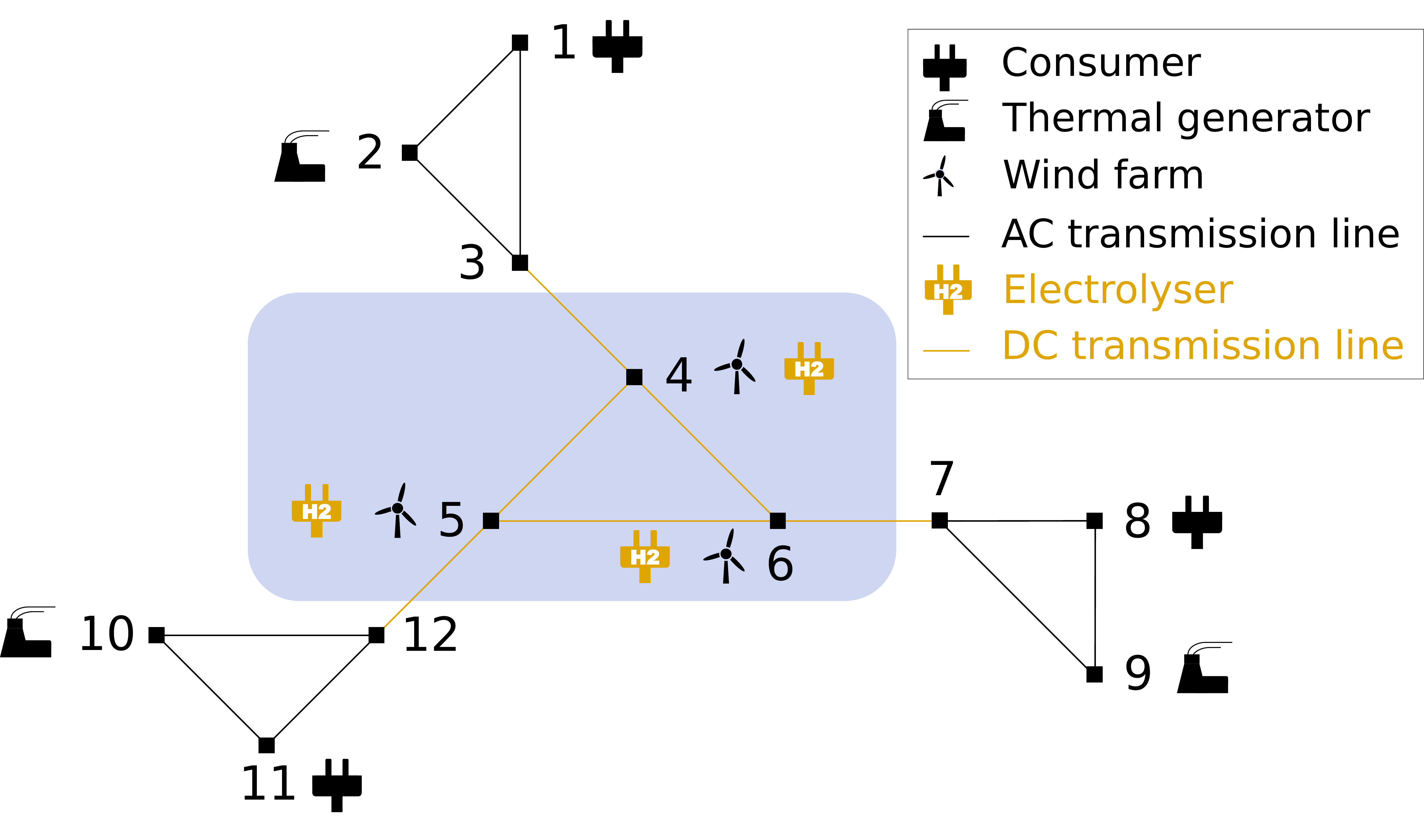}
    \caption{Electricity grid. The symbols indicate the location of generators, consumers and transmission lines with a black color for existing elements (AC transmission lines, thermal generation units, consumers and offshore wind farms) and a yellow color for elements of which the capacity is a decision variable (DC transmission lines and electrolyzers).}
    \label{fig:network}
\end{figure}

We consider three unit electrolyzer investment cost levels (low, medium and high) to illustrate the possible role of offshore electrolysis under each of the four electricity market designs in Tab. \ref{tab:marketdesigns}.

\begin{figure}
     \centering
     \begin{subfigure}[b]{0.5\textwidth}
         \centering
         \includegraphics[width=\textwidth]{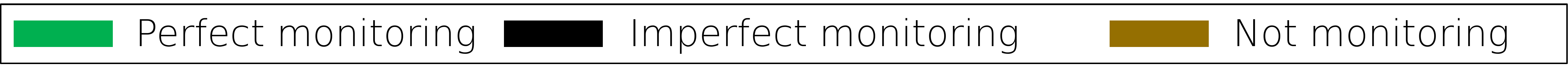}
     \end{subfigure}
     \hfill
     \begin{subfigure}[b]{0.115\textwidth}
         \centering
         \includegraphics[width=\textwidth]{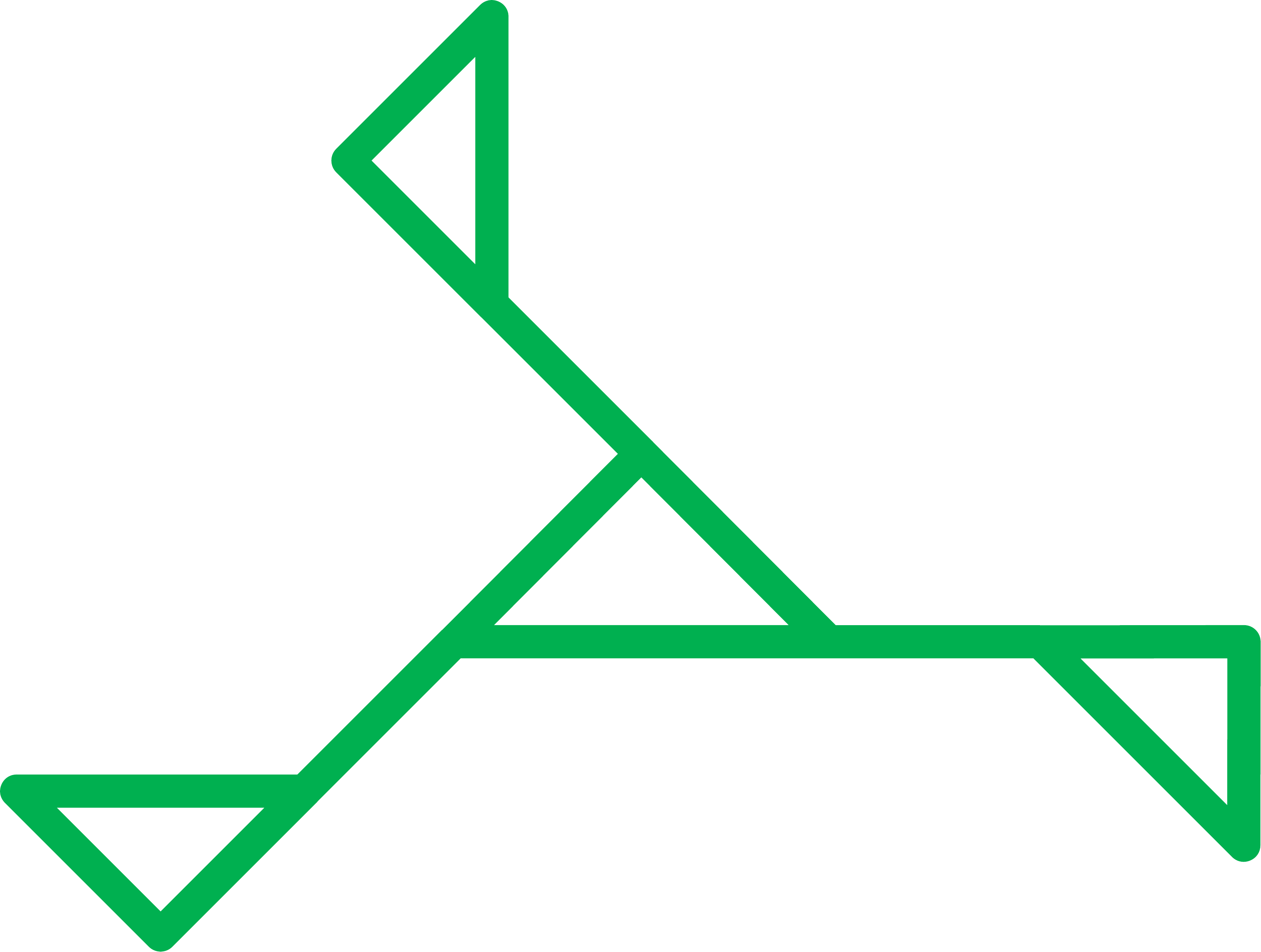}
         \caption{FNP}
         \label{fig:gridconstraints1}
     \end{subfigure}
     \hfill
     \begin{subfigure}[b]{0.115\textwidth}
         \centering
         \includegraphics[width=\textwidth]{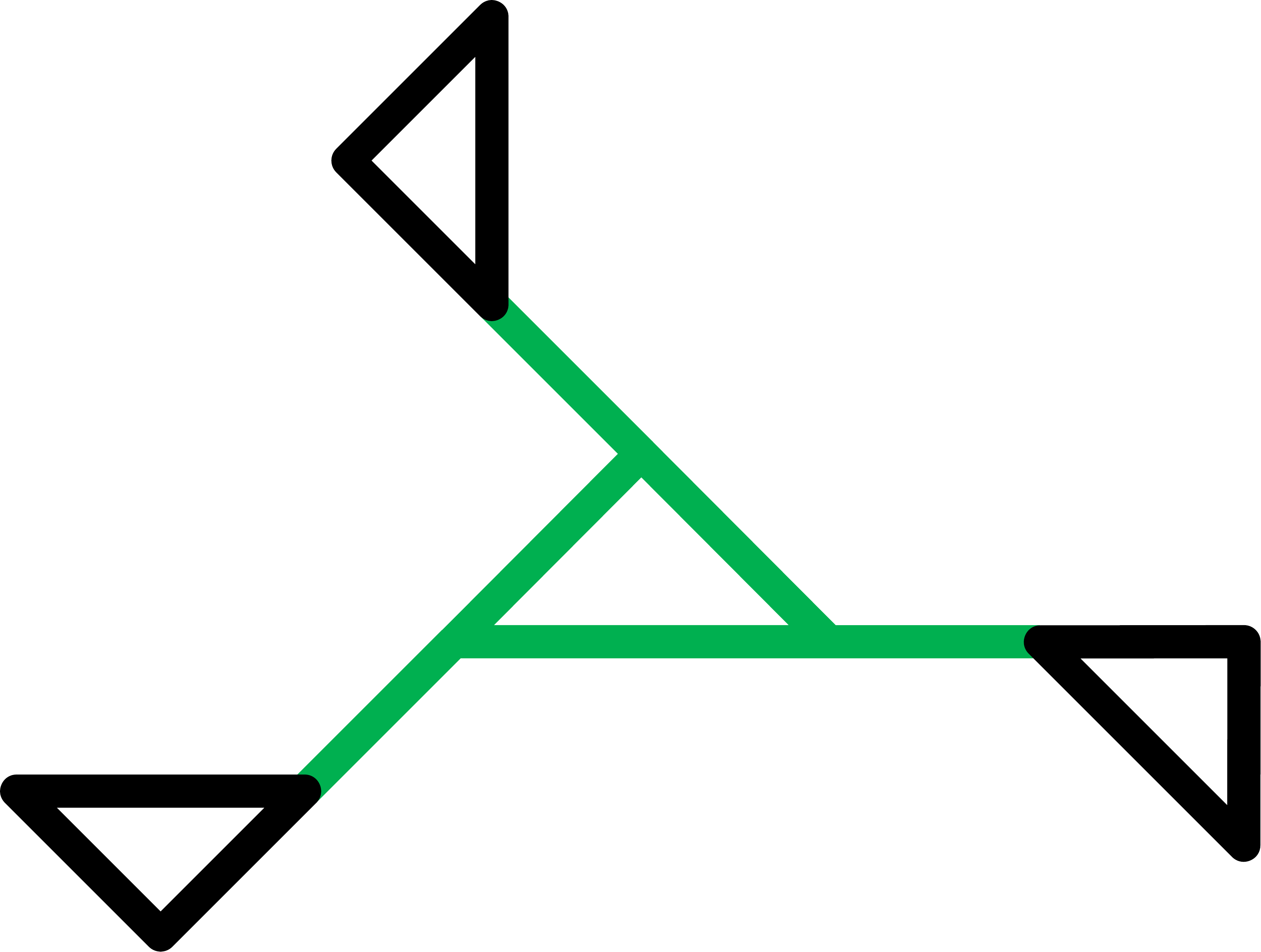}
         \caption{ONP}
         \label{fig:gridconstraints2}
     \end{subfigure}
     \hfill
     \begin{subfigure}[b]{0.115\textwidth}
         \centering
         \includegraphics[width=\textwidth]{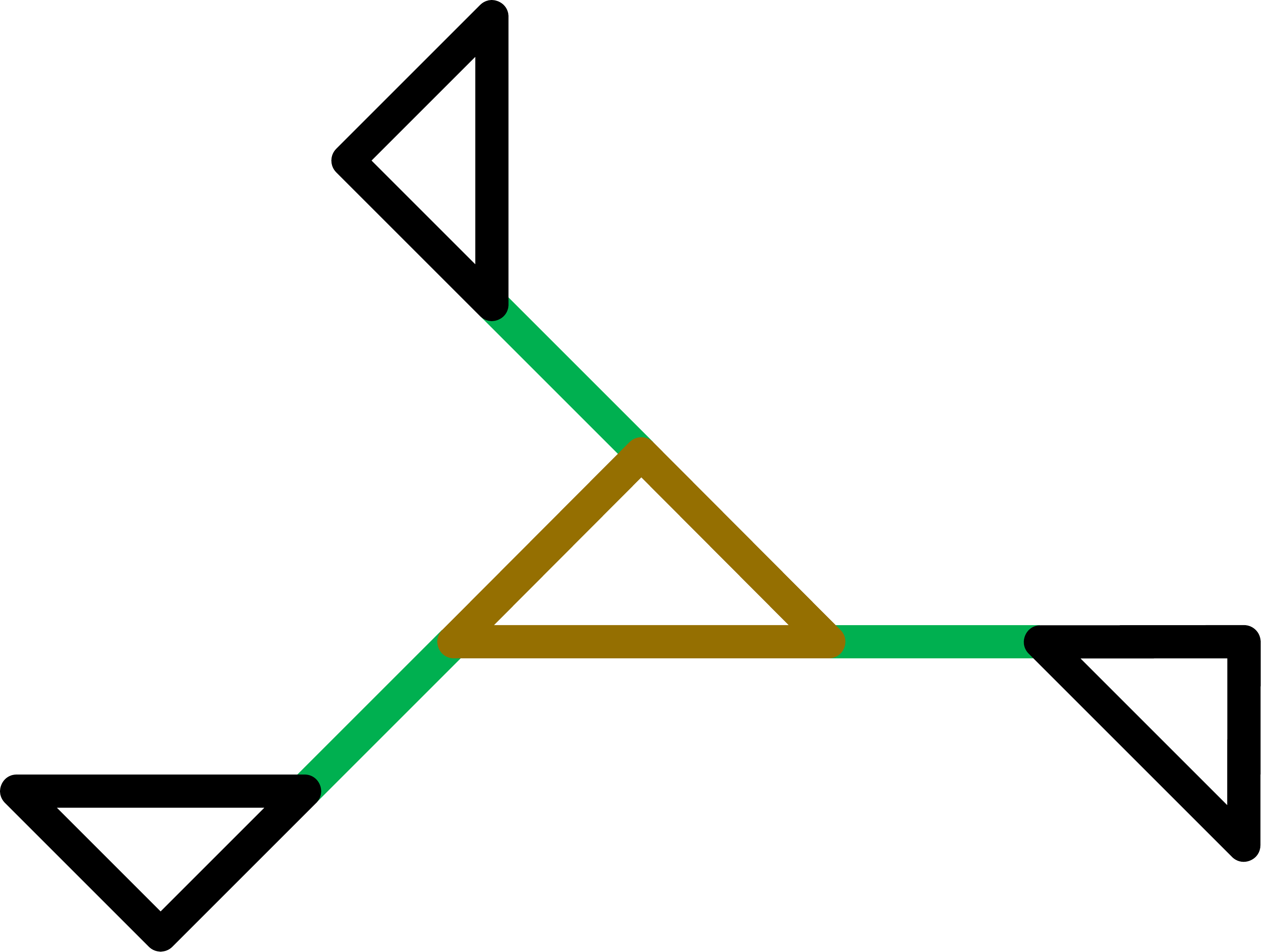}
         \caption{OZP}
         \label{fig:gridconstraints3}
     \end{subfigure}
     \hfill
     \begin{subfigure}[b]{0.115\textwidth}
         \centering
         \includegraphics[width=\textwidth]{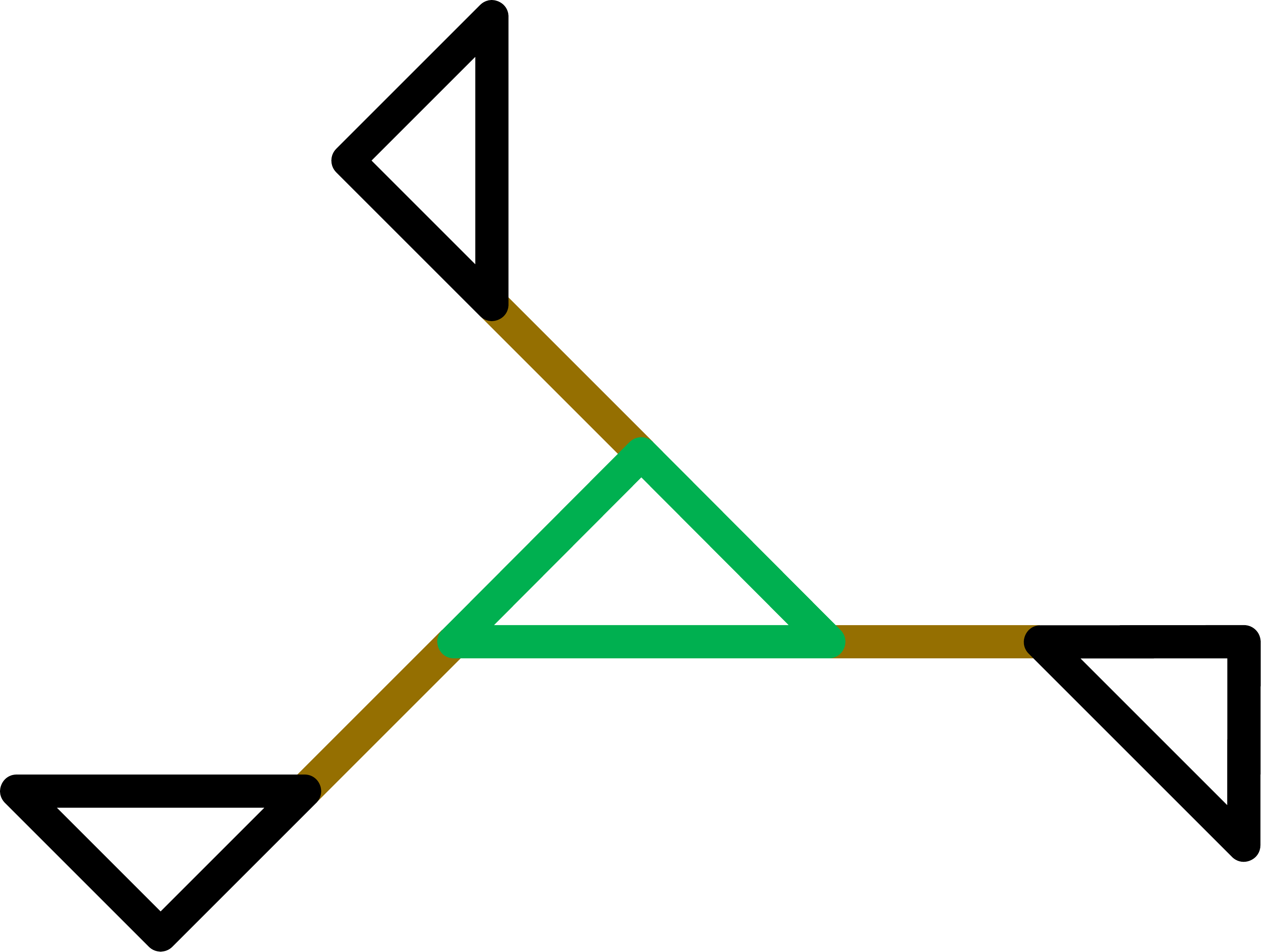}
         \caption{FZP}
         \label{fig:gridconstraints4}
     \end{subfigure}
        \caption{Accuracy of monitoring the flows on the AC and DC transmission lines in the market clearing algorithm under the four different pricing methods.}
        \label{fig:gridconstraints}
\end{figure}

Figure \ref{fig:gridconstraints} shows how each pricing method implies a different monitoring of the flows on the transmission lines in the market clearing. While FNP perfectly monitors all flows, the other pricing methods involve (partly) zonal pricing following FBMC. Therefore, the flows on the AC lines ($\mathcal{L}$) are imperfectly monitored using zonal PTDFs under ONP, OZP and FZP. The DC-lines follow the NTC approach and are only monitored if they are cross-border lines ($\mathcal{H}'$).

\subsection{Results} \label{sec:results}

We discuss market designs implications on (i) the overall welfare, (ii) the profitability of offshore wind farms and (iii) offshore electricity prices. 

\subsubsection{Implications on welfare} \label{sec:results_welfare}

Figure \ref{fig:welfare} presents the market surplus $\sum_t S_t$, aggregated over all time steps and normalised w.r.t. the market surplus under FNP and the lowest electrolyzer unit investment cost. The difference between the market surplus and overall welfare (Eq. (\ref{eq:UL_obj})) consists of the redispatch cost $\sum_t R_t$, the electrolyzer investment cost $GIC$, and the transmission investment cost $TIC$.

We make three observations. Firstly, ONP and OZP outperform FZP in terms of welfare despite a higher market surplus and a lower spending on electrolyzer and transmission capacities under FZP. Specifically, the market surplus under FZP amounts to 97.9\%, 96.4\% and 96.3\% of the reference market surplus for the three investment costs of electrolyzers, while it ranges from 87.7\% to 96.3\% under ONP and from 88.7\% to 96.6\% under OZP. The market surplus is higher under FZP compared to ONP and OZP because more grid constraints are considered in the day-ahead market under the latter two cases. Besides, the combined investments in electrolyzer and transmission capacities are consistently lower under FZP (0.7\%, 1.5\% and 0.1\% of the reference market surplus) compared to other market designs. As a consequence, the difference in overall welfare between ONP and OZP, on the one hand, and FZP, on the other hand, is entirely driven by lower redispatch costs $R_t$ due to a better grid representation in the market clearing problem (see Fig. \ref{fig:gridconstraints}). The redispatch cost under FZP is 20.9\%, 23.3\% and 25.4\% of the reference market surplus for the three investment costs of electrolyzers, while it is 9.7\% at most under other market designs. 

\begin{figure}[h]
    \centering
    \includegraphics[width=\textwidth,width=0.5\textwidth]{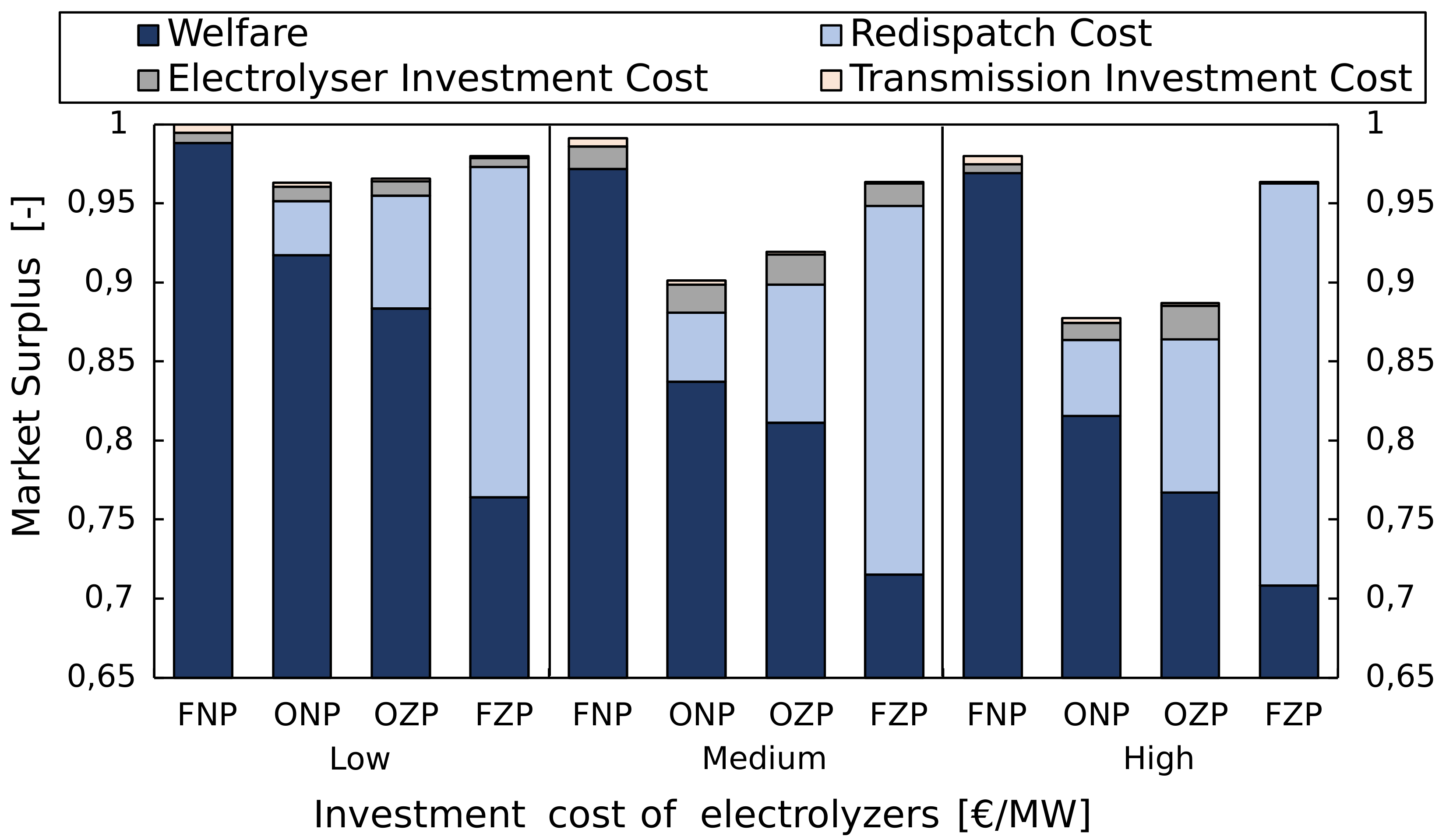}
    \caption{Market surplus $\sum_t S_t$, normalised w.r.t. the market surplus under nodal pricing and lowest electrolyzer investment cost. The overall welfare (Eq. (\ref{eq:UL_obj})) is indicated in blue.}
    \label{fig:welfare}
\end{figure}
\begin{figure}[h]
    \centering
    \includegraphics[width=0.5\textwidth]{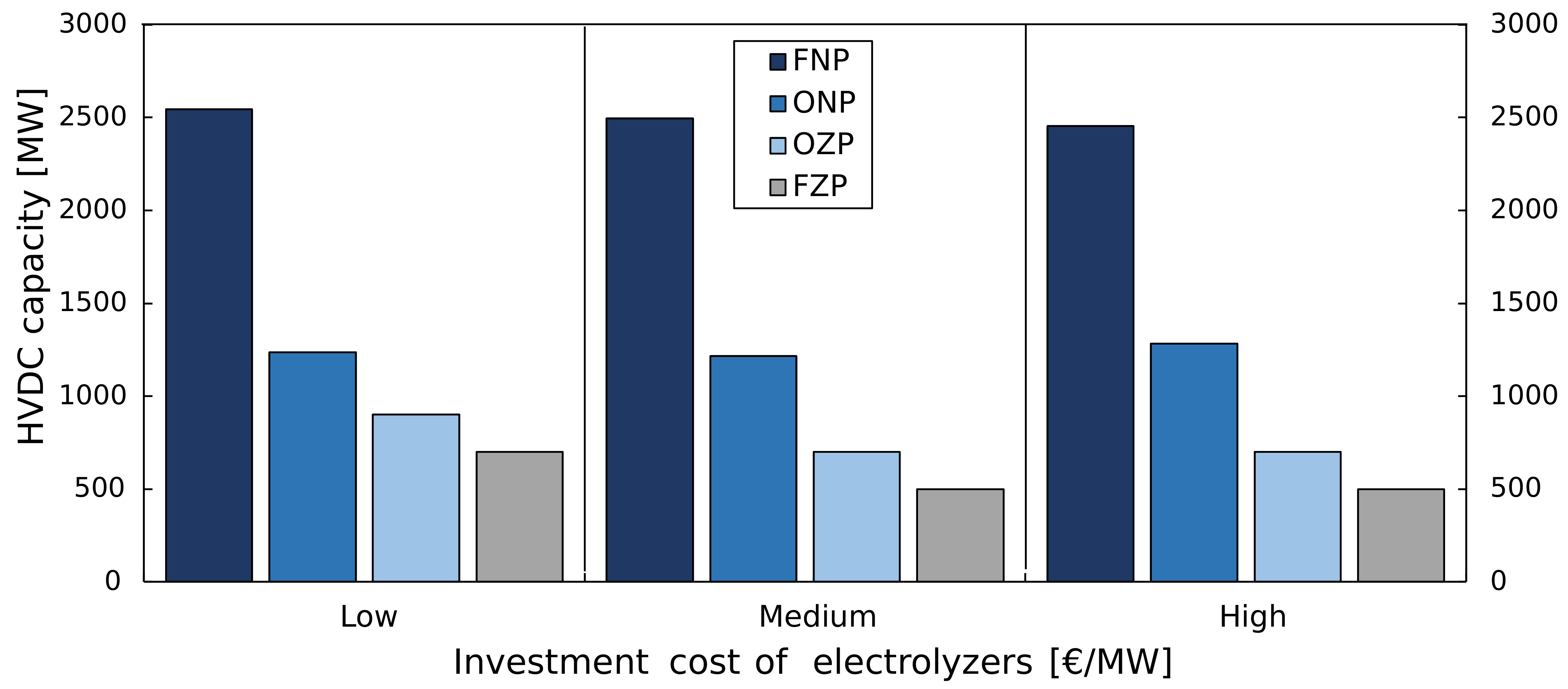}
    \caption{HVDC capacity.}
    \label{fig:HVDC}
\end{figure}
\begin{figure}[!h]
    \centering
    \includegraphics[width=0.5\textwidth]{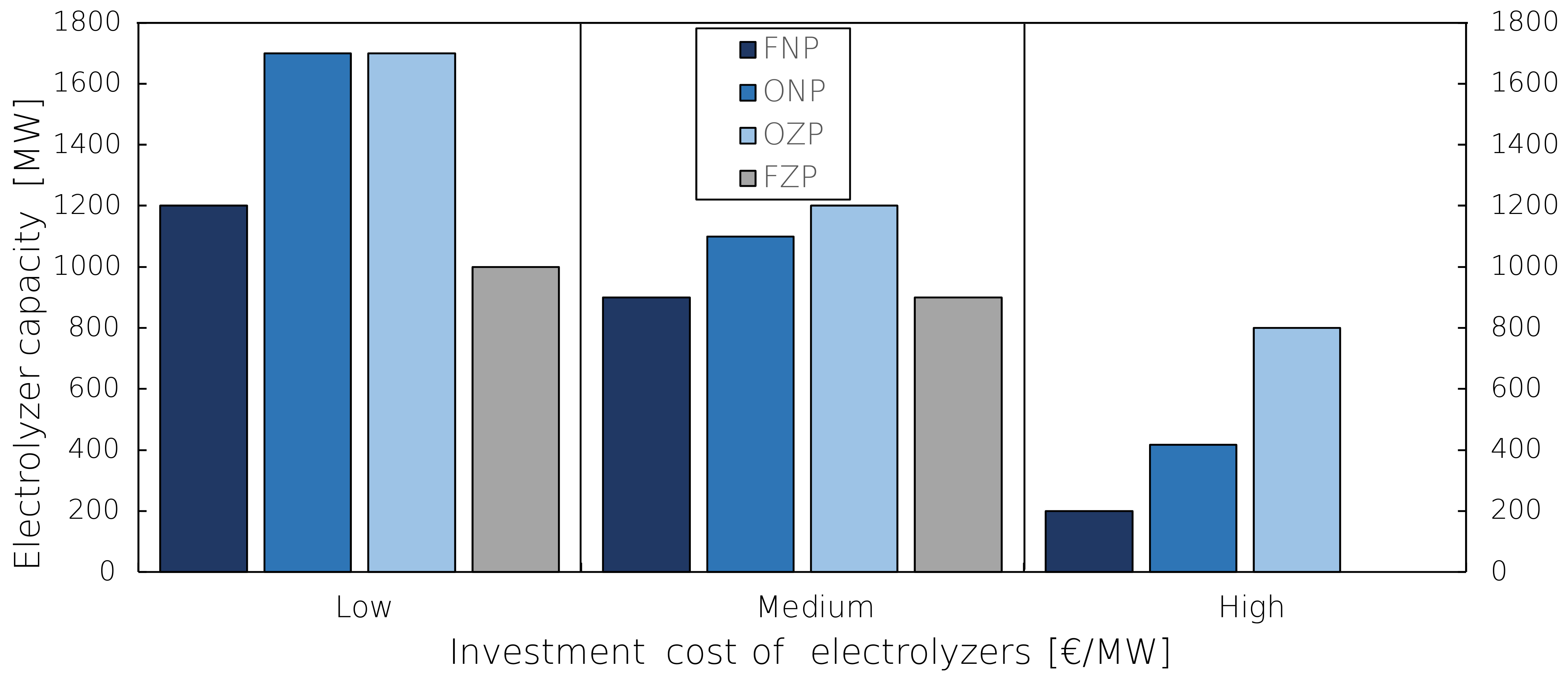}
    \caption{Electrolyzer capacity.}
    \label{fig:electrolyzer}
\end{figure}

\textcolor{blue}{A more accurate representation of the transmission constraints in the market clearing algorithm allows capturing the contribution of potential HVDC investments to increased welfare and reduced redispatch costs. Recall that intra-zonal HVDC links are not considered in the market clearing problem under OZP or FZP. Under ONP, onshore zonal pricing restricts flows on the offshore network elements, limiting the attainable wellfare gains (see also Section \ref{sec:results_prices}).} In our case study a more accurate grid representation results in more HVDC capacity (Fig. \ref{fig:HVDC}) while a market clearing problem with the most crude grid representation relies on the possibility for redispatch to avoid congestion. The HVDC capacity under ONP, OZP and FZP is lower because the additional market surplus as a result of installing more HVDC capacity is limited under ONP, OZP and FZP compared to FNP. It does, however, lower the redispatch cost.

Secondly, welfare increases with a decrease in the unit investment cost of electrolyzers regardless of the market design. This is because a higher electrolyzer capacity, which is shown in Fig. \ref{fig:electrolyzer}, allows generating a larger market surplus, outweighing the larger electrolyzer investment cost. The market surplus grows with 2.1\% (FNP), 8.6\% (ONP), 8.0\% (OZP) and 1.6\% (FZP) of the reference market surplus while the change in combined spending in electrolyzers and HVDC lines is less than 1.4\% under all market designs. In addition, a higher electrolyzer capacity allows reducing the redispatch cost, which drops with 1.3\% (ONP), 2.6\% (OZP) and 4.4\% (FZP) of the reference market surplus. 

Thirdly, the welfare gain associated with a lower unit investment cost of the electrolyzers is larger under ONP (10.1\% of the reference market surplus) and OZP (11.6\% of the reference market surplus) than under FNP or FZP. Lower offshore wholesale prices under ONP and OZP compared to the other market designs trigger higher investments in electrolyzers (see further) and thus an increasing impact of their investment costs on the total welfare. Hence, the electrolyzer capacity is higher under FNP, OZP and ONP than under FZP.\footnote{\textcolor{blue}{Note that the bilevel modeling approach inherently allows multiple optimal solutions to the lower level. The upper level problem may "exploit" this multiplicity of solutions and select the lower level solution that maximizes the upper level objective. Similarly, this characteristic of bilevel models implies that while the reported welfare and HVDC investment are optimal, this may be achieved through different electrolyzer capacity investments, day-ahead market clearing and redispatch actions.}}

\subsubsection{Implications on wind farm \textcolor{blue}{\& electrolyzer} profitability} \label{sec:results_profitability}

Figure \ref{fig:profit} presents the operational profit for all offshore wind farms. Firstly, the profit is the highest under FZP closely followed by FNP, while profits are significantly lower under ONP and particularly OZP. For example, in case of a low investment cost of electrolyzers, the profit amounts to 1.72 M\EUR{} under FZP, followed by 1.66 M\EUR{} (FNP), 1.34 M\EUR{} (ONP) and 0.60 M\EUR{} (OZP). Secondly, the profit increases with a decrease in investment cost of electrolyzers regardless of the market design. Specifically, it grows with 0.10 M\EUR{} (FNP), 0.79 M\EUR{} (ONP), 0.32 M\EUR{} (OZP) and 0.15 M\EUR{} (FZP) when the investment cost decreases from high to low. Thirdly, the profit increase is the largest under ONP with 0.79 M\EUR{}. The drivers behind these effects are curtailment and day-ahead offshore electricity prices, which we discuss in the following.

Figure \ref{fig:curtailment} presents the wind energy curtailment as part of the market clearing and the redispatch actions, expressed as a fraction of the available offshore wind energy. Curtailment under FZP is at least 6 times higher than under other other market designs. This is a consequence of the crude representation of grid constraints in the market as the majority of the curtailment under FZP occurs as part of the redispatch actions (see also redispatch costs in Fig. \ref{fig:welfare}). The presence of electrolyzers limits the need for curtailment as it serves as an electricity consumer at the same node at which wind power is generated. Specifically, we find a negative correlation of -0.52 between wind power curtailment and electricity consumption by electrolyzers.

\begin{figure}
    \centering
    \includegraphics[width=0.5\textwidth]{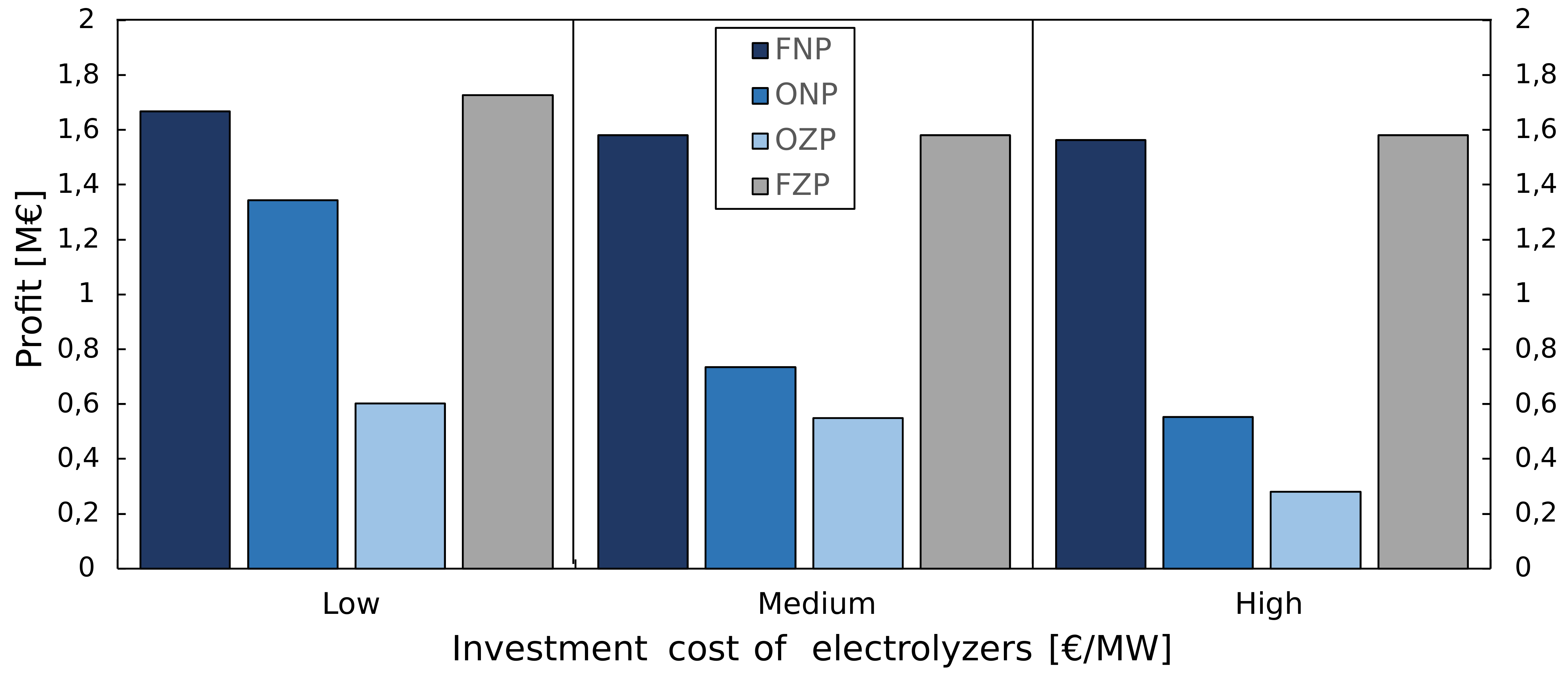}
    \caption{Operational profit of the offshore wind farms.}
    \label{fig:profit}
\end{figure}

\begin{figure}
    \centering
    \includegraphics[width=0.5\textwidth]{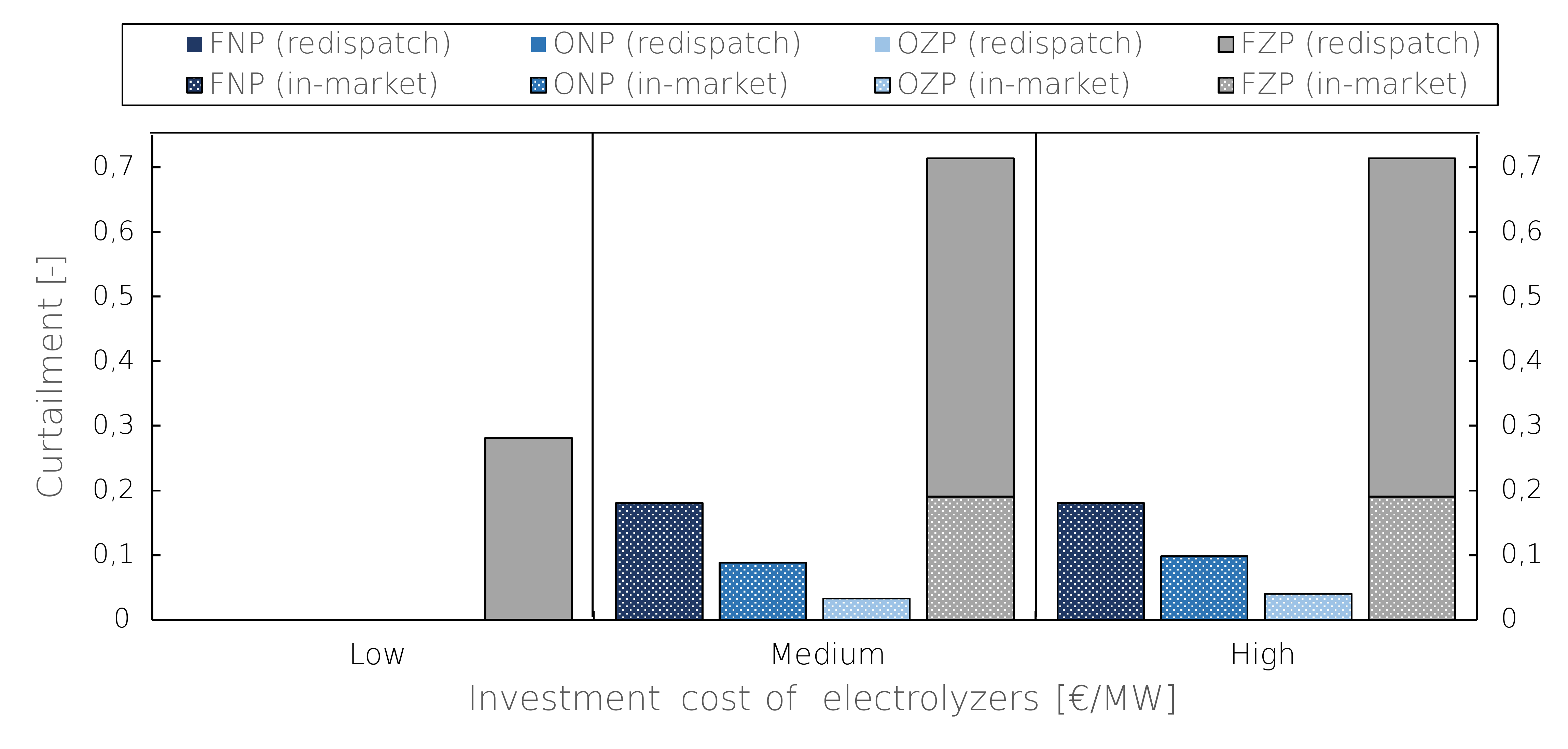}
    \caption{Curtailment of offshore wind energy, expressed as a fraction of the available offshore wind energy. We distinguish between curtailment as part of the market clearing and the redispatch actions.}
    \label{fig:curtailment}
\end{figure}

\textcolor{blue}{By design, the electrolyzers do not secure a net profit unless the optimal electrolyzer capacity equals the assumed upper limit. In case an electrolyzer's capacity equals this upper bound, its operating profits can outweigh its investment cost indicating that a higher electrolyzer capacity would increase social welfare. If the electrolyser capacity does not equal its upper limit, operating profits offset the investment costs and the electrolysers' net profit is zero.}

\subsubsection{Implications on offshore electricity prices} \label{sec:results_prices}

Figure \ref{fig:price} presents the day-ahead electricity prices for offshore wind farms (node 4, 5 and 6). Offshore electricity prices are, averaged over time and offshore nodes, the highest under FZP. Specifically, the average price amounts to 71.5 \EUR{}/MWh under FZP compared to 65.2 \EUR{}/MWh under FNP at a low investment cost, to 69.3 \EUR{}/MWh under FZP compared to 64 \EUR{}/MWh under FNP at a medium investment cost and to 69.3 \EUR{}/MWh under FZP compared to 62.7 \EUR{}/MWh under FNP at a high investment cost. Average prices are the highest under FZP because the flow on HVDC lines between an offshore and an onshore node (3-4, 6-7 and 5-12) is not monitored nor constrained in the market clearing problem. Hence, price convergence with the onshore zones occurs by construction. Contrary, the flows on all HVDC lines between an offshore and an onshore node (3-4, 6-7 and 5-12) are monitored and constrained in the market clearing problem under FNP, ONP and OZP because they serve as cross-zonal HVDC lines. As a result, electricity prices in the offshore zone(s) are zero whenever the commercial transmission capacity restricts flows, lowering the average offshore prices under FNP, ONP and OZP compared to FZP.

\begin{figure*}
    \centering
    \begin{tikzpicture}
        \node[inner sep=0pt] (russell) at (0,0)
        {\includegraphics[width=\textwidth]{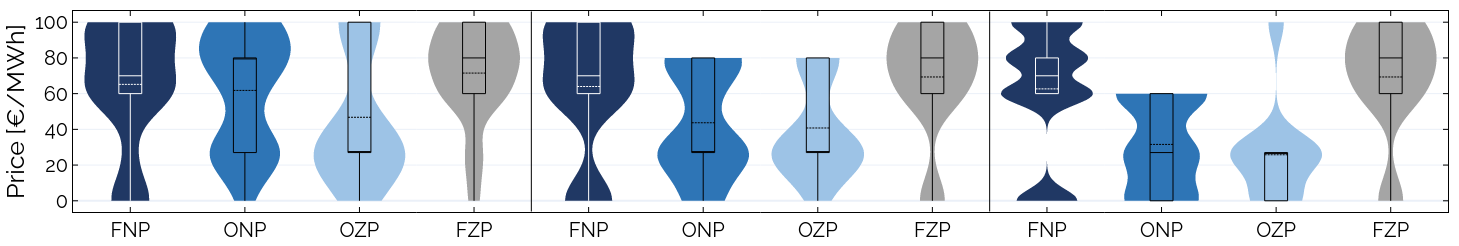}}; %,height=7.2cm
        % \draw[black] (-7.8,-2.6) rectangle (-2.6,-2.1);
        \node at (-5.2,-1.7){\small{Low}};
        % \draw[black] (-2.2,-2.6) rectangle (3.15,-2.1);
        \node at (0.5,-1.7){\small{Medium}};
        % \draw[black] (3.55,-2.6) rectangle (8.75,-2.1);
        \node at (6.1,-1.7){\small{High}};
        \node at (0.5,-2.05){\small{Investment cost of electrolyzers [€/MW]}};
    \end{tikzpicture}
  \caption{Prices for offshore wind farms. The boxes show the 25th and 75th quantiles and the median while the dashed line shows the average.}
  \label{fig:price}
\end{figure*}
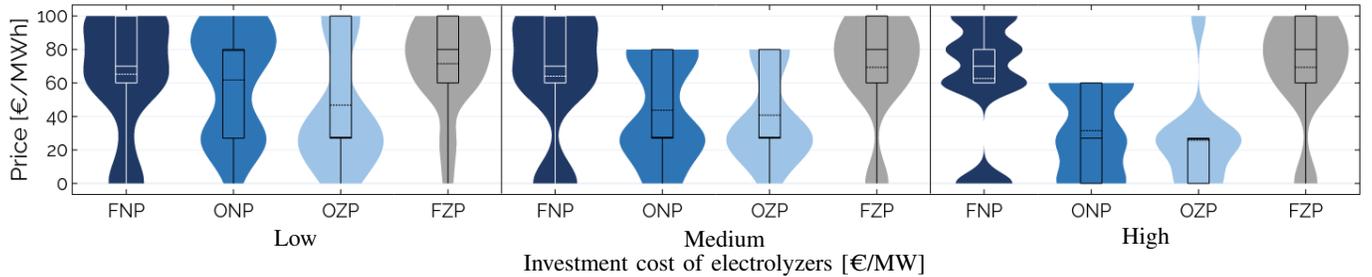

Despite that the flows on all HVDC lines between an offshore and an onshore node (3-4, 6-7 and 5-12) are monitored and constrained in the market clearing problem under FNP, ONP and OZP, offshore prices are different between these three pricing methods. Specifically, the average price is consistently higher under FNP (65.2 \EUR{}/MWh, 64 \EUR{}/MWh and 62.7 \EUR{}/MWh) than under ONP (61.8 \EUR{}/MWh, 43.7 \EUR{}/MWh and 31.6 \EUR{}/MWh) which in turn is higher than under OZP (46.7 \EUR{}/MWh, 40.8 \EUR{}/MWh and 25.8 \EUR{}/MWh). As expected, the electrolyzer capacity is inversely correlated with the average electricity price per offshore node and timestep. 

We identify two drivers explaining the price differences between FNP, ONP and OZP: (i) the (optimised) HVDC transmission capacity, and (ii) not considering offshore internal grid constraints under OZP. Firstly, under FNP, the flows on both AC and DC lines are perfectly monitored and limited. Contrary, under ONP and OZP, the flows on the onshore AC lines are imperfectly monitored and limited due to the characteristics of a flow-based market clearing. The zonal character of the onshore market implies that the flows on the offshore DC lines are restricted more than under FNP (see Eq. (\ref{eq:LL1_zonal_con1})). This leads to less value for HVDC transmission capacity (see Fig. \ref{fig:HVDC}). Hence, there is less price convergence between the offshore nodes and the onshore zones. Secondly, the average offshore price is lower under OZP than under ONP because, under OZP, congestion between an offshore node and the connecting onshore node immediately results in a price difference between all three offshore nodes and the onshore zone. This drives down the average offshore price because the marginal cost of offshore wind farms is assumed to be zero. 

A lower unit investment cost of electrolyzers consistently drives up the average offshore price under all market designs. This is the result of the combination of two drivers: (i) because thermal generation units produce more electricity and therefore can set the price more often, and (ii) because the willingness-to-pay of electrolyzers $U^{H2}_{e,t}$ (27 \EUR{}/MWh in this case study) sets the price more often. The increase of the average offshore price with decreasing investment costs of electrolyzers is the largest under ONP and OZP. Specifically, it increases with 96\% (from 31.6 \EUR{}/MWh to 61.8 \EUR{}/MWh) under ONP and 81.4\% (from 25.8 \EUR{}/MWh to 46.8 \EUR{}/MWh) under OZP compared to 4\% (from 62.7 \EUR{}/MWh to 65.2 \EUR{}/MWh) under FNP and 3\% (from 69.3 \EUR{}/MWh to 71.5 \EUR{}/MWh) under FZP. This is because the electrolyzers drive up the offshore electricity price such that the level of price convergence (see Fig. \ref{tab:uniqueprices}) increases the most under these market designs, which is reflected in the market surplus that also witnesses the largest increase under ONP and OZP with decreasing investment costs of electrolyzers. As a result, the presence of a consumer at an offshore node in the form of an electrolyzer can contribute to the profitability of offshore wind farms.

Finally, the offshore HVDC capacity impacts the level of electricity prices (\ref{fig:price}) and price convergence. Table \ref{tab:uniqueprices} shows the percentage of time during which a specific amount of unique prices exists. One unique price implies full price convergence across all nodes/zones. The largest amount of hours in which there are one or two unique prices consistently occurs under FNP compared to ONP, OZP and FZP. Specifically, this is during 80\%, 86.7\% and 76.7\% of time for the three levels of investment costs of electrolyzers compared to at most 20\% under any other market design. This is contra-intuitive because FNP allows for up to 12 unique prices. However, FNP signals scarcity of transmission capacity better than any other market design leading to the most HVDC deployment, hence, the highest level of price convergence.

\begin{table*}
    \centering
    \caption{Percentage of time during which a specific amount of unique prices exists. One unique price implies full price convergence across all nodes/zones.}
    \begin{tabular}{l | l l l l | l l l l | l l l l }
    \toprule
        & \multicolumn{4} {c|} {Low} &  \multicolumn{4} {c|} {Medium} & \multicolumn{4} {c} {High} \\
        & FNP & ONP & OZP & FZP & FNP & ONP & OZP & FZP & FNP & ONP & OZP & FZP \\
    \midrule
        1 unique price & 6.7\% & 3.3\% & 0\% & 0\% & 20\% & 0\% & 0\% & 0\% & 20\% & 0\% & 0\% & 0\%\\
        2 unique prices & 73.3\% & 16.7\% & 0\% & 6.7\% & 66.7\% & 0\% & 0\% & 20\% & 56.7\% & 0\% & 0\% & 20\%\\
        3 unique prices &  10\% & 60\% & 90\% & 93.3\% & 13.3\% & 43.3\% & 80\% & 80\% & 23.3\% & 13.3\% & 50\% & 80\%\\
        4 unique prices & 10\% & 20\% & 10\% & - & 0\% & 46.7\% & 20\% & - & 0\% & 70\% & 50\% & -\\
        5 unique prices & 0\% & 0\% & - & - & 0\% & 10\% & - & - & 0\% & 16.7\% & - & -\\
    \bottomrule
    \end{tabular}
    \label{tab:uniqueprices}
\end{table*}

\section{Discussion} \label{sec:discussion}

Our results show that FNP is superior to ONP, OZP and FZP in terms of economic welfare. Yet, there exist political hurdles that prevent the implementation of FNP in many regions \cite{eicke}. Instead, FZP is the active pricing methodology in many regions including, e.g., the EU, UK, Brazil and Australia. \textcolor{blue}{Many of these hurdles are rooted in concerns around the redistributional impacts for existing consumers and generators. In an offshore context, there are no or very few existing generators/loads. We would therefore expect less opposition against ONP or OZP, which our case study reveals to be valuable alternatives to FZP.} Specifically, despite leading to a lower market surplus and an increased spending on HVDC transmission capacity and electrolyzer capacity, these market designs increase overall welfare by significantly decreasing offshore wind energy curtailment and redispatch costs, originating either offshore or onshore. The latter is the result of a more accurate representation of grid constraints in the market clearing algorithm under ONP and OZP compared to FZP. \textcolor{blue}{Offshore wind developers remain skeptical of ONP and OZP as they fear lower average electricity prices, as well as higher price and volume risks due to congestion. Our results indeed show tat} the overall increase in welfare comes at the cost of a lower profitability of existing offshore wind farms. The lower profitability is explained by a lower average offshore price as congestion between onshore and offshore nodes is signaled via price differences with a certain spatial granularity. \textcolor{blue}{This effect is stronger under OZP compared to ONP.} 

\textcolor{blue}{Technically, the challenges with zonal and nodal offshore market design are similar, and often related to the interaction with other markets, such as the balancing market. With little demand and no "back-up" generators offshore, system operators structurally have to rely on balancing services provided by neighboring zones or nodes. Notably, the controllability of HVDC technology with regard to power flows offers Transmission System Operators (TSOs) more flexibility in managing grid constraints, potentially easing some technical challenges.}

A different market design than FZP would also impact public support mechanisms for offshore wind farms like, e.g., Contract-for-Differences (CfDs). Specifically, the cost for governments would increase under CfDs assuming ceteris paribus. Under FZP, support for offshore wind developers is implicit in the high (socialized) redispatching costs, i.e., wind developers receive artificially high grid-agnostic day-ahead electricity prices but may be curtailed in real-time. Alternative to increased support for offshore wind farms, our results suggest that implementing nodal pricing onshore in addition to ONP, which is identical to FNP, also brings the profitability of wind farms close to levels under FZP.

Offshore consumers like, e.g., offshore electrolyzers can serve as a catalyst to decrease the remaining welfare gap between FNP, on the one hand, and ONP and OZP on the other hand. A lower offshore average price under ONP and OZP compared to FNP could allow offshore electrolyzers to enter the market, which in turn increases the market surplus and reduces redispatch costs. Besides, it reduces wind energy curtailment and increases the offshore average price, leading to an increased profitability of offshore wind farms. Therefore, there is an important interaction between support mechanisms for offshore wind and potential support mechanisms for electrolyzers. The latter would reduce the need for the former because of the increasing offshore prices with more electrolyzers, especially under ONP and OZP.

\section{Conclusion} \label{sec:conclusion}

Different strategies exist, or are part of the ongoing debate, to integrate offshore energy hubs in wholesale electricity markets. We develop a bilevel optimization model to simulate the implications of four market designs: full nodal pricing, offshore nodal pricing, offshore zonal pricing, and full zonal pricing. The model captures four temporal stages: (i) offshore HVDC transmission investment, (ii) offshore electrolyzer investment, (iii) wholesale electricity market clearing, and (iv) redispatch actions. Offshore wind generation investment is exogenously determined. 

We find that full nodal pricing is superior to offshore nodal, offshore zonal and full zonal pricing. Yet, there exist political hurdles that prevent the implementation of full nodal pricing in many regions. Our case study reveals that offshore nodal and offshore zonal pricing can serve as valuable alternatives to full zonal pricing. Specifically, these market designs could increase welfare as a consequence of decreased offshore wind energy curtailment and decreased redispatch costs despite a lower market surplus. However, the welfare gain comes at the cost of a lower profitability of existing offshore wind farms due to congestion between onshore and offshore nodes. The latter benefits the business case for offshore consumers like, e.g., offshore electrolyzers, which, in turn, can have a positive effect on the offshore average price. Therefore, if proven profitable in real world power systems, offshore electrolyzers can serve as a catalyst to decrease the remaining welfare gap between full nodal pricing, on the one hand, and offshore nodal and offshore zonal pricing on the other hand. Policy makers and regulators should, hence, carefully coordinate electricity market design with policies supporting the development of offshore energy infrastructure.

\textcolor{blue}{Future work could focus on how balancing mechanisms should be set-up in under offshore nodal and zonal or full nodal pricing schemes, in which most balancing energy would need to be provided by onshore assets in different bidding zones. Focal points in this discussion are the interaction with day-ahead electricity markets and transmission capacity allocation, the impact on offshore wind farms and the use of offshore assets such electrolyzers as a source of balancing energy.}

\bibliography{IEEEabrv,references}
\bibliographystyle{IEEEtran}

\section*{Appendices}

\subsection*{Appendix A: Parameters in the case study}

We assume that each thermal generation unit and each offshore wind farm has an installed capacity $\bar{y}_g$ of 2000 MW and 1000 MW respectively. The operational cost $MC_g$ of each thermal generation unit is 60 \EUR{}/MWh at node 2, 80 \EUR{}/MWh at node 9 and 100 \EUR{}/MWh at node 10, while we assume it is zero for the offshore wind farms. Figure \ref{fig:loadfactor} presents the assumed availability factor $LF_{g,t}$ at each offshore node as well as the price inelastic demand $\bar{D}_{n,t}$ at each node and time step. There are six peaks in both available wind energy and demand. However, both do not always peak at the same time steps. We assume that the utility to consume electricity as end product $U_{n}$ is 200 \EUR{}/MWh and to produce hydrogen $U^{H2}_{e,t}$ is 27 \EUR/MWh. The penalisation parameter in the redispatch problem $a$ is 1000. 

We assume that each AC line has a thermal capacity of 700 MW and all have the same impedance. Both the RAMs and NTCs are set identical to the thermal capacity of the considered line. The zonal PTDFs are derived from the nodal PTDFs by assuming values for GSKs that are pro rata with the generation capacity in a zone \cite{van2016flow}.

\textcolor{blue}{We assume three levels for the unit investment cost of electrolyzers $C_e$: low (50 \EUR{}/MW), medium (150 \EUR{}/MW) and high (250 \EUR{}/MW).} The maximal electrolyzer capacity $\bar{D}^{H2}_{e,t}$ is 600 MW per offshore node. The investment cost of a DC line $C_h$ is 20 \EUR{}/MW. The maximal capacity of a new DC line $\bar{F}^{DC}_h$ is 2000 MW. 

\begin{figure}[h]
    \centering
    \includegraphics[width=0.5\textwidth]{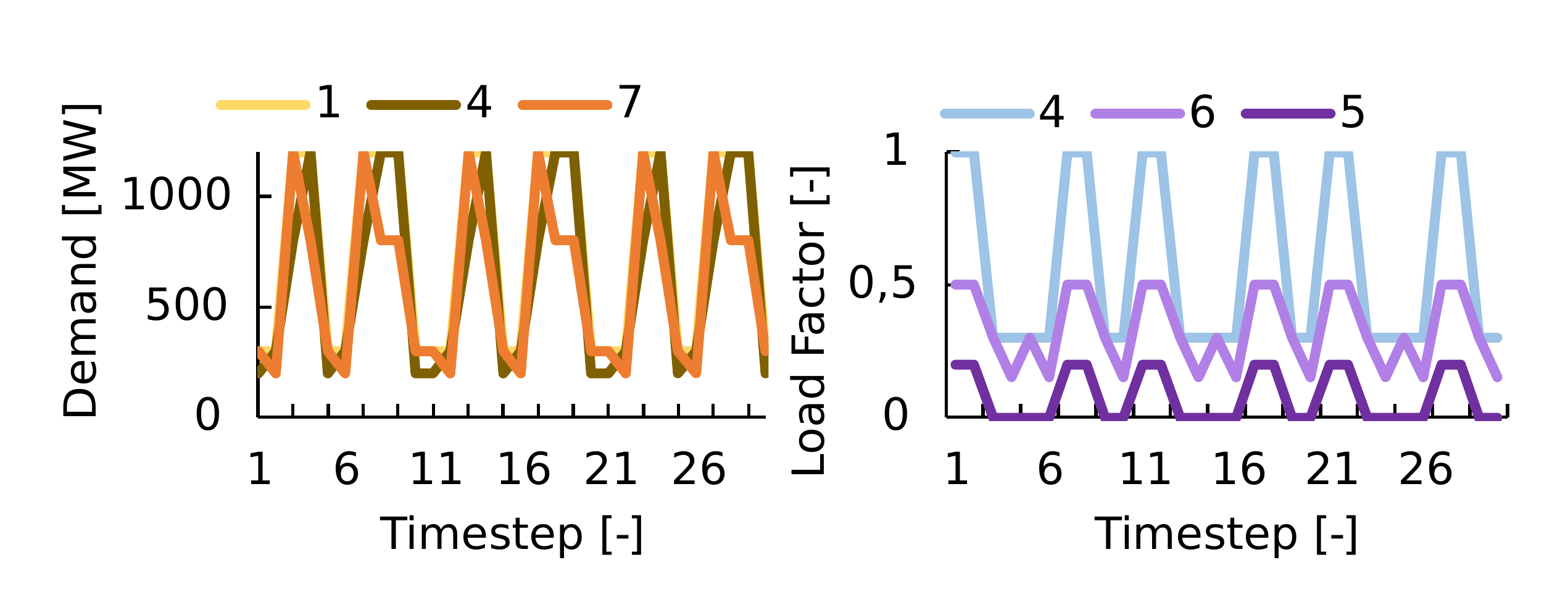}
    \caption{Availability factors of the offshore wind farms and price inelastic demand at each node and time step.}
    \label{fig:loadfactor}
\end{figure}

\endgroup
\end{document}